\documentclass[%
twocolumn,
superscriptaddress,
notitlepage,
nofootinbib,
floatfix
]{revtex4-1}

\pdfoutput=1

\usepackage{amsmath}
\usepackage{amsfonts}
\usepackage{amssymb}
\usepackage{mathrsfs}
\usepackage{amsthm}
\usepackage{wasysym}
\usepackage[usenames]{color}
\usepackage[usenames]{xcolor}
\usepackage[unicode,colorlinks=true,linkcolor=blue,citecolor=blue,pdfusetitle]{hyperref}
\usepackage[all]{hypcap}
\usepackage{graphicx}
\usepackage{xspace}
\usepackage{verbatim}
\usepackage{enumerate}
\usepackage{mathrsfs}
\usepackage{cleveref}

\usepackage[T1]{fontenc}
\usepackage[utf8]{inputenc}
\usepackage{txfonts}
\usepackage{microtype}

\usepackage{multirow}

\newcommand{\be}{\begin{equation}}
\newcommand{\ee}{\end{equation}}
\newcommand{\bs}{\begin{split}}
\newcommand{\es}{\end{split}}

\newcommand{\cD}{\mathscr{D}}
\newcommand{\cDD}{\mathscr{D}^\dagger}
\newcommand{\nn}{\nonumber}
\newcommand{\lrbrk}[1]{\left(#1\right)}
\newcommand{\lrsbrk}[1]{\left[#1\right]}

\newcommand{\lm}{\ell m}

\newcommand{\lmo}{\ell m \omega}
\newcommand{\lpmo}{\ell^\prime m \omega}
\newcommand{\lpmos}{\ell^\prime m \omega^*}

\newcommand{\lmmmo}{\ell \,\text{-}m\, \text{-}\omega}
\newcommand{\lpmmmo}{\ell^\prime \,\text{-}m\, \text{-}\omega}
\newcommand{\lpmmmos}{\ell^\prime \,\text{-}m\, \text{-}\omega^*}
\newcommand{\lmmmos}{\ell \,\text{-}m\, \text{-}\omega^*}

\newcommand{\tbar}{\bar{t\vphantom{\theta}}}

\newcommand{\comments}[1]{}

\begin{document}
\title{Tidal response and near-horizon boundary conditions for spinning exotic compact objects}

\author{Baoyi Chen}
\email{baoyi@tapir.caltech.edu}
\affiliation{Walter Burke Institute for Theoretical Physics and Theoretical Astrophysics, M/C 350-17, California Institute of Technology, Pasadena, CA 91125, US}

\author{Qingwen Wang}
\email{qwang@perimeterinstitute.ca}
\affiliation{Perimeter Institute and the University of Waterloo, 31 Caroline St N, Waterloo, ON N2L 2Y5, Canada}

\author{Yanbei Chen}
\email{yanbei@tapir.caltech.edu}
\affiliation{Walter Burke Institute for Theoretical Physics and Theoretical Astrophysics, M/C 350-17, California Institute of Technology, Pasadena, CA 91125, US}

\date{\today}

\begin{abstract} 
Teukolsky equations for $|s|=2$ provide efficient ways to solve for curvature perturbations around Kerr black holes.  Imposing regularity conditions on these perturbations on the future (past)  horizon corresponds to imposing an in-going (out-going) wave boundary condition. 
For exotic compact objects (ECOs) with external Kerr space time, however, it is not yet clear how to physically impose boundary conditions for curvature perturbations on their boundaries.   We address this problem using the Membrane Paradigm, by considering a family of zero-angular-momentum fiducial observers (FIDOs)  that float right above the horizon of a linearly perturbed Kerr black hole.  
From the reference frame of these observers, the ECO will experience tidal perturbations due to in-going gravitational waves, respond to these waves, and generate out-going waves. As it also turns out, if both in-going and out-going waves exist near the horizon, the Newman Penrose (NP) quantity $\psi_0$ will be numerically dominated by the in-going wave, while the NP quantity $\psi_4$ will be dominated by the out-going wave --- even though both quantities contain full information regarding the wave field.   In this way, we obtain the ECO boundary condition in the form of a relation between $\psi_0$ and the complex conjugate of $\psi_4$, in a way that is determined by the ECO's tidal response in the FIDO frame.  We explore several ways to modify gravitational-wave dispersion in the FIDO frame, and deduce the corresponding ECO boundary condition for Teukolsky functions.  Using the Starobinsky-Teukolsky identity, we subsequently obtain the boundary condition for $\psi_4$ alone, as well as for the Sasaki-Nakamura and Detweiler's functions.    As it also turns out, reflection of spinning ECOs will generically mix between different $\ell$ components of the perturbations fields, and be different for perturbations with different parities. 
It is straightforward to apply our boundary condition to computing gravitational-wave echoes from spinning ECOs, and to solve for the spinning ECOs' quasi-normal modes. 
\end{abstract}

\maketitle

\section{Introduction}

A black hole (BH) is characterized by the event horizon, a boundary of the space-time region within which the future null infinity cannot be reached.  
The existence of a horizon has lead to the simplicity of black-hole solutions in general relativity and modified theories of gravity, although the notion of a horizon has also led to technical and conceptual problems. 
First of all, at the classical level, the even horizon has a {\it teleological nature}: its shape at a particular time-slice of a spacetime depends on what happens to the future of that slide.  Even if we are provided with a full numerical solution of the Einstein's equation (e.g., in the form of all metric components in a particular coordinate system), it is much harder to determine the location of the event horizon than {\it trapped surfaces}, whose definitions are more local.

In classical general relativity, it has been shown that a singularity (or singularities) should always exist inside the event horizon~\cite{Penrose:1969pc}, this requires that quantum gravity must be used to study the space-time geometry inside black holes.  Naively, one expects corrections when space-time curvature is at the Planck scale.  However, the unique causal structure of the horizon already leads to non-trivial quantum effects, e.g., Hawking radiation~\cite{Hawking:1974sw,Hawking:1976ra}. From considerations of quantum gravity, it has been proposed that space-time geometry near the horizon can be modified, even at scales larger than the Planck scale~\cite{Lunin:2001jy,Lunin:2001jy,Mathur:2005zp,PrescodWeinstein:2009mp,Braunstein:2009my,Almheiri:2012rt,Bianchi:2020miz,Bianchi:2020bxa,Giddings:2017jts}.    It has also been conjectured that a phase transition might occur during the formation of black holes, leading to non-singular, yet extremely compact objects~\cite{Mazur:2004fk,Mathur:2008nj}.  All these considerations (or speculations) lead to a similar class of objects: their external space-time geometries mimic those of black holes except very close to the horizon.  We shall refer to these objects as {\it Exotic Compact Objects} (ECOs).

Followed by the unprecedented discovery of gravitational waves from the binary BH merger event GW150914~\cite{Abbott:2016blz}, and follow-up observations of an order of $\sim 100$ binary black-hole merger events~\cite{LIGOScientific:2018mvr,Abbott:2020niy},  we now know that dark compact objects do exist in our universe, and that their space-time geometry and dynamics are consistent with those of black holes in general relativity, better than order unity, and at scales comparable to the sizes of the black holes.  Observations by the Event Horizon Telescope (EHT) provides yet another avenue toward near-horizon physics of black holes~\cite{Akiyama:2019cqa,Akiyama:2019brx,Akiyama:2019sww,Akiyama:2019bqs,Akiyama:2019fyp,Akiyama:2019eap,Gralla:2019xty,Giddings:2019jwy}.

Since the horizon is defined as the boundary of the unreachable region hence ``absorbs'' all radiation, instead of asking whether the horizon exists, a more testable  question might be how absorptive the horizon is: any potential modifications to classical general relativity near the surface of an ECO, be it quantum or not, may impose a different physical boundary condition near the horizon.   That is, for any incoming gravitational radiations, they not only can fall into the dark object, but may also get reflected, and then propagate to the infinity.  In the context of a point particle orbiting a black-hole candidate, this was studied as a modified tidal interaction~\cite{Fang:2005qq,Li:2007qu,Datta:2019epe}.   Alternatively, a stronger probe of the reflectivity is provided by waves that propagate toward the horizon of the final (remnant) black hole after the merger of two black holes --- in the  form of repeated GW echoes at late times in the ringdown signal of a binary merger event~\cite{Cardoso:2016rao,Cardoso:2016oxy,Cardoso:2017cqb,Mark:2017dnq,Abedi:2016hgu,Micchi:2020gqy,Micchi:2019yze,Conklin:2019smy,Sago:2020avw,Burgess:2018pmm}.  Following this line of thought, the gravitational echoes has been extensively studied in different models of near-horizon structures~\cite{Conklin:2017lwb,Oshita:2018fqu,Wang:2019rcf,Cardoso:2019apo,Buoninfante:2020tfb,Bueno:2017hyj}.  Even though the idea of ECOs might be speculative, one can always regard the search for ECOs as one to quantify the darkness of the final objects in binary merger events,  and in this way its importance cannot be overstated.

A major missing piece of the current echo program is how to apply boundary condition near the horizon for curvature perturbations obtained from the Teukolsky equation.  This was discussed, by Nakano et al.~\cite{Oshita:2019sat} and Wang and Afshordi~\cite{Wang:2019rcf}, but for Kerr there are still more details to fill in --- even though Kerr echoes have already been studied by several authors~\cite{Micchi:2020gqy,Micchi:2019yze,Conklin:2019smy,Sago:2020avw,Burgess:2018pmm}.  This is the main problem we would like to address in this paper. 

Imposing a near-horizon boundary condition is more straightforward in Schwarzschild spacetime.  The Schwarzschild metric perturbations can be fully constructed from solutions of the  Regge-Wheeler equation~\cite{Regge:1957td} and the Zerilli equation~\cite{Zerilli:1971wd}, both of which are wave equations that have regular asymptotic behaviors at horizon and infinity.  These metric perturbations can then be used to connect to the response of the ECO to external perturbations.   In the Kerr spacetime, perturbations are most efficiently described by the $s=\pm 2$ Teukolsky equations~\cite{Teukolsky:1973ha} for curvature components that are projected along null directions, and therefore are less directly connected to tidal perturbations and responses of an ECO.  Furthermore, the Teukolsky equations for the $s= \pm 2$ cases do not have short-range potentials, and result in solutions that do not have the standard form of incoming and out-going waves, leading to certain difficulties in finding numerical solutions.  
 
 To solve the second issue,   the Teukolsky equations can be transformed into wave-like equations with short-ranged potentials, namely the Sasaki-Nakamura (SN) equations, via  the Chandrasekhar-Sasaki-Nakamura (CSN) transformation~\cite{Sasaki:1981kj,Sasaki:1981sx,Hughes:2000pf}. In order to define the near-horizon reflection of waves in the Kerr spacetime, it was proposed that the reflection should be applied to the SN functions --- as  has been widely used in many literatures regarding gravitational wave echoes~\cite{Conklin:2017lwb,Nakano:2017fvh,Sago:2020avw,Li:2007qu,Du:2018cmp,Micchi:2020gqy,Micchi:2019yze,Conklin:2019smy,Sago:2020avw}. 
 Despite the short-rangedness of the SN equation, the physical meaning of SN functions are less clear than the Teukolsky functions, especially in the Kerr case.

For the Kerr spacetime, Thorne, Price and MacDonald introduced the ``Membrane Paradigm'' (MP)~\cite{Thorne:1986iy}, by considering a family of fiducial observers (FIDOs) with zero angular momentum.  World lines of the collection of these observers form a ``membrane'', which can be used as a proxy to think about the interaction between the black hole and the external universe.   In order to recover the pure darkness of the black hole, the membrane must have the correct complex (in fact purely resistive) impedance for each type of flux/current, so that nothing is reflected.  For example, the membrane must have the correct specific viscocity in order for gravitational waves not to be reflected, and the correct (electric) resistivity in order for electromagnetic waves not to be reflected.    Extensive discussions were made on the physics viewed by the FIDOs, in particular tidal tensors measured by these observers in presence of gravitational waves.   The picture was more recently used to visualize space-time geometry using Tendex and Vortex picture~\cite{Nichols:2011pu,Zhang:2012jj,Nichols:2012jn}. 

It has been proposed that reflectivity of ECOs can be modeled by altering the impedance of the ECO surface~\cite{Maggio:2020jml,Wang:2019rcf,Oshita:2019sat}.  In this paper, we generalize this point of view to ECOs with nonzero spins.  It is worth mentioning that the the membrane paradigm point of view has been taken by Datta {\it et al.}~\cite{Datta:2019epe,Datta:2020rvo} to study the  tidal heating of Kerr-like ECOs, although reflection of waves by the ECO was not described.   In this paper we shall continue along with the membrane paradigm, and propose a physical definition of the ECO's reflectivity.

In order to do so, we make a careful connection between Teukolsky functions, which efficiently describe wave propagation between the near-horizon region and infinity, and in-going and out-going tidal waves in the FIDO frame of the Membrane Paradigm. 
We then obtain boundary conditions for the Teukolsky equations  in terms of tidal responses of the ECO in the FIDO frame.  Here the {\it fundamental assumption} that we rely upon --- as has also been made implicitly in previous ECO reflectivity literatures ---  is that the ECO has a simple structure in the FIDO frame --- for example as a distribution of exotic matter that modifies dispersion relation of gravitational waves in the FIDO frame.

We organize the paper as follows. 
In Sec.~\ref{sec:near-horizon-BC}, by considering individual FIDOs, we introduce the modified boundary conditions in Teukolsky equations based on the tidal response of the ECO, and obtain input-output relations for Teukolsky equations in terms of that tidal response. 
In Sec.~\ref{sec:R-in-emergent-G}, we more specifically consider a Rindler coordinate system near the horizon, and put our discussion into a more firm ground by relating the Teukolsky functions to Riemann tensor components in this coordinate system.  We further consider modified gravitational-wave dispersion relations in the Rindler frame, and relate these relations to the ECO's tidal response. 
In Sec.~\ref{sec:boundary-cond-func}, we translate our reflection model into a model which fits most literatures on gravitational-wave echoes, in particular making connections to the SN formalism. 
In Sec.~\ref{sec:waveform-qnm}, we apply our method to obtain the echo waveform as well as the quasi-normal modes (QNMs) of the ECOs.  Showing that even- and odd-parity waves will generate different echoes, and generalize the breaking of  QNM {\it isospectrality} found by Maggio {\it et al.}~\cite{Maggio:2020jml} to the spinning case.   
In Sec.~\ref{sec:summary}, we summarize all results and propose possible future works.

{\it Notation.}~We choose the natural units $G=c=1$, and set the black hole mass $M=1$.  The following symbols are also used throughout the paper: 
\begin{align}
\Delta & =r^2-2 r + a^2\,,  \\
\Sigma &= r^2 + a^2 \cos^2\theta\,, \\
\rho &= - \lrbrk{r-i a \cos\theta}^{-1}\,.
 \end{align} 
Here $(t,r,\theta,\phi)$ are the Boyer-Lindquist coordinates for Kerr black holes and $a$ is the black hole spin. The Kerr horizons are at the Boyer-Lindquist radius $r_H = 1 + \sqrt{1-a^2}$, while the inner horizons are at $r_C = 1 - \sqrt{1-a^2}$.
The angular velocity of the horizon is given by $\Omega_H = a/\lrbrk{2 r_H}$.
The tortoise coordinate is defined by
\begin{equation}
r_* = r + \frac{2 r_H}{r_H - r_C}\ln\lrbrk{\frac{r-r_H}{2}} - \frac{2 r_C}{r_H - r_C}\ln\lrbrk{\frac{r-r_C}{2}}\,.
\end{equation}

\section{The reflection boundary condition from tidal response}
\label{sec:near-horizon-BC}

In a $(3+1)$-splitting of the spacetime, the Weyl curvature tensor $C_{abcd}$ naturally gets split into an ``electric'' part, which is responsible for the {\it tidal effect}, and a ``magnetic'' part, which is responsible for the {\it frame-dragging effect}.  From now on, we will focus on the electric part, as it gives rise to the gravitational stretching and squeezing, i.e. the tidal force, which drives the geodesic deviations of particles that are  {\it slowly moving with respect to that slicing}. 

\begin{figure}
\includegraphics[width=0.35\textwidth]{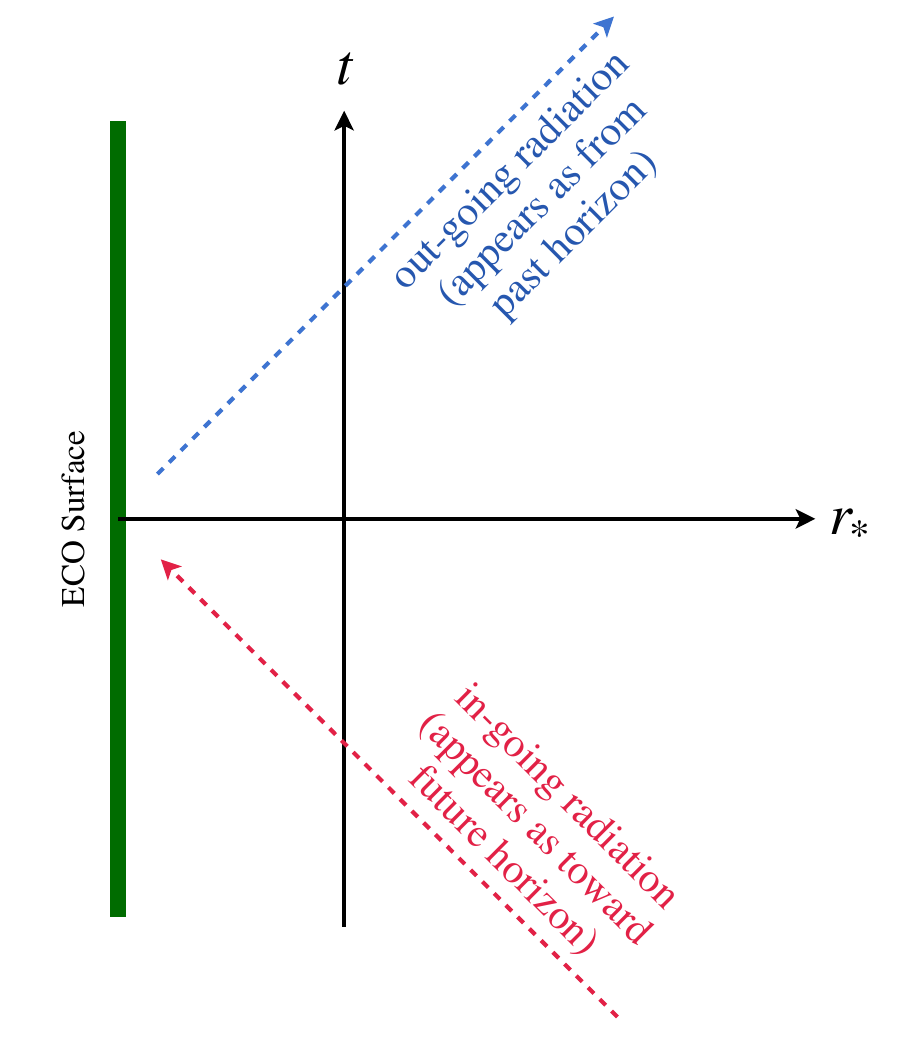}
\caption{Waves that propagate toward the ECO surface can be approximated as propagating toward the future horizon, while those originate from the ECO surface can be approximated as originating from the past horizon.\label{fig:futurepast}}
\end{figure}

In MP, a relation is established between the Newman-Penrose quantity $\psi_0$ near the future horizon and components of the tidal tensors in the FIDO frame.   In this section, we will extend this to include waves ``originating from past horizon'', which really are waves in the vicinity of the horizon but propagate toward the positive $r$ direction, see Fig.~\ref{fig:futurepast}.   More specifically, we seek to derive the relation among the tidal tensor components, the incoming waves, and the ``reflected'' (outgoing) waves due to the tidal response.  This will establish our model of near-horizon reflection for the Teukolsky equations.

\begin{figure}
\includegraphics[width=0.30\textwidth]{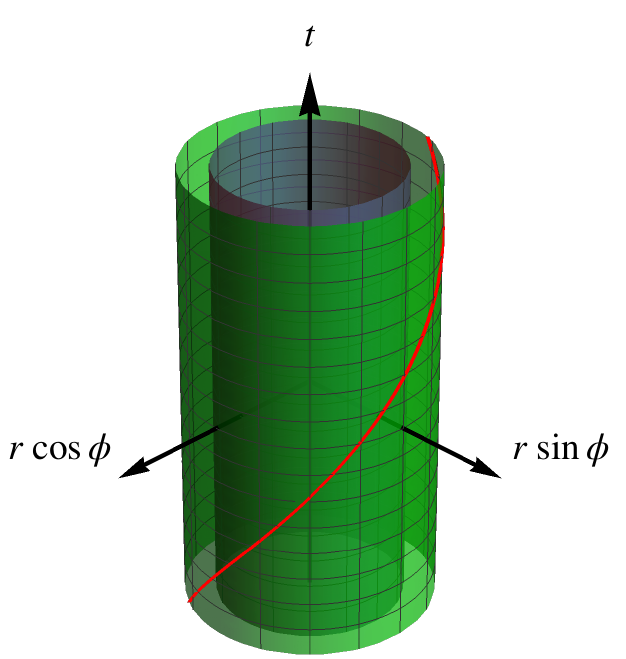}
\caption{Trajectory of the FIDO in a constant $\theta$ slice of the Kerr spacetime in the $(t,r\cos\phi,r\sin\phi)$ coordinate system.  Here the green surface indicates the ECO surface with $r=b$, while the black surface indicates the Kerr horizon. Each FIDO has $r=b$, but has $(t,\phi) = (t,\phi_0+\Omega_H t)$. \label{fig:surface}}
\end{figure}

\subsection{FIDOs}
Starting from the Boyer-Lindquist coordinate system $(t,r,\theta,\phi$),  FIDOs in the MP are characterized by constant $r$ and $\theta$, but $\phi = {\rm const} + \omega_\phi t $, with 
\begin{equation}
\omega_\phi = \frac{2 a r}{\Xi} \,.
\end{equation}
and 
\begin{equation}
\Xi = (r^2+a^2)^2 - a^2\Delta \sin^2\theta\,.
\end{equation}
Each FIDO carries an orthonormal tetrad of~\footnote{Note that MP uses different notations for the $\rho$, $\Sigma$.}
\begin{align}
\label{eq:fido-tetrad}
\vec{e}_{\hat r} & =\sqrt{\frac{\Delta}{\Sigma}} \vec{\partial}_r\,,\;
&\vec{e}_{\hat \theta}& =\frac{\vec{\partial}_\theta}{\sqrt{\Sigma}}\,,\; \\ \nn
\vec{e}_{\hat \phi}& = \sqrt{\frac{\Sigma}{\Xi}}\frac{\vec{\partial}_\phi}{\sin\theta}\,,\;
&\vec{e}_{\hat 0}& = \frac{1}{\alpha}\lrbrk{\vec{\partial}_t+\omega_\phi \vec{\partial}_\phi}\,,
\end{align}
with 
\begin{equation}
\alpha  =\sqrt{\frac{\Sigma \Delta}{\Xi}}\,.
\end{equation}
Here $\vec e_{\hat 0}$ is the four-velocity of the FIDO.  The FIDOs have  zero angular momentum (hence are also known as Zero Angular-Momentum Observers, or ZAMOs),  since $\vec e_{\hat 0}$ has zero inner product with $\vec\partial_\phi$.  Here $\alpha$ is called the redshift factor,  also known as the lapse function, since it relates the proper time of the FIDOs and the coordinate time $t$.

 Near the horizon, we have  $\alpha\rightarrow 0$; FIDO's tetrads are related to  the Kinnersly tetrad~\cite{Kinnersley:1969zza} via
\begin{align}
\label{eq:fido-kinnersky}
\vec l \approx \sqrt{\frac{\Sigma}{\Delta}} (\vec{e}_{\hat 0} +\vec{e}_{\hat r} )\,, 
\vec n \approx \sqrt{\frac{\Delta}{\Sigma}} \frac{\vec{e}_{\hat 0} -\vec{e}_{\hat r} }{2}\,,
\vec m \approx \frac{e^{i\beta}(\vec{e}_{\hat \theta} +i\vec{e}_{\hat \phi})}{\sqrt{2}}\,,
\end{align}
where 
\begin{equation}
 \beta = - \tan^{-1} \lrbrk{\frac{a \cos\theta}{r_H}} \,.
\end{equation}

\subsection{Tidal Tensor Components}

Let us now introduce the {\it electric-type} tidal tensor $\mathcal{E}$ as viewed by FIDOs, which can be formally defined as~\cite{Zhang:2012jj}
\begin{equation} 
\mathcal{E}_{ij} = {h_i}^a {h_j}^c C_{abcd} U^b U^d \,.
\end{equation}
Here $U = \vec{e}_{\hat 0}$ is the four-velocity of FIDOs as in Eq.~\eqref{eq:fido-tetrad}, and ${h_{i}}^a = {\delta_i}^a + U_i U^a$ is the projection operator onto the spatial hypersurface orthogonal to $U$.
In particular, we look at the $mm$-component of the tidal tensor, as the gravitational-wave stretching and squeezing will be along these directions.  Near the horizon, the tidal tensor component is then given by
\begin{equation}
\label{eq:Emm-to-psi}
\mathcal{E}_{mm} = C_{\hat 0 m\hat 0 m} \approx  -\frac{\Delta}{4\Sigma}\psi_0 - \frac{\Sigma}{\Delta} \psi_4^* \,.
\end{equation}

For convenience, let us define a new variable ${}_{s}\Upsilon$ which is the solution to the Teukolsky equation with spin weight $s$. For $s=\pm2$ we have
\begin{align}
&{}_{-2}\Upsilon \equiv \rho^{-4}\psi_4 \,, & {}_{+2}\Upsilon \equiv \psi_0 &\,.
\end{align}
We briefly review the Teukolsky formalism in Appendix.~\ref{app:Teukolsky-eq}. For perturbations that satisfy the linearized vacuum Einstein's equation (in this case the Teukolsky equation), at $r_* \rightarrow -\infty$, in general we can decompose ${}_s\Upsilon$ using the spin-weighted spheroidal harmonics ${}_{s}S_{\lmo}(\theta)$, and write
\begin{align}
\label{eq:Teukolsky-H-decompose}
{}_{-2}\Upsilon (t, r_*, \theta, \phi) =&\sum_{\lm}\int \frac{d \omega}{2\pi }  e^{-i\omega t} 
\,{}_{-2}S_{\lmo}(\theta)\,e^{im\phi}  \\ \nn
&\times \lrsbrk{Z^{\rm hole}_{\lmo} \Delta^2e^{-ikr_*} + Z^{\rm refl}_{\lmo} \, e^{ikr_*}} \,, \\ 
\label{eq:Teukolsky2-H-decompose}
{}_{+2}\Upsilon (t, r_*, \theta, \phi) =&\sum_{\lm}  \int \frac{d \omega}{2\pi }e^{-i\omega t} 
\,{}_{+2}S_{\lmo}(\theta)\,e^{im\phi}  \\  \nn
 & \times \lrsbrk{ Y^{\rm hole}_{\lmo} \, \Delta^{-2} e^{-ikr_*} + Y^{\rm refl}_{\lmo} \, e^{ikr_*} } \,,
\end{align} 
where $k \equiv \omega \, - \,m \, \Omega_{H}$.  We use the shorthand $\sum_{\lm} \equiv \sum_{\ell=2}^{\infty} \sum_{m=-\ell}^{\ell}$, in which $\ell$ is the multipolar index, and $m$ is the azimuthal quantum number.  Note this $m$ here should not be confused with the label $m$ in the Kinnersly tetrad basis.  Here $Z_{\lmo}$ and $Y_{\lmo}$ are amplitudes for the radial modes, with ``hole'' labeling the left-propagation modes into the compact object (in this paper, left means toward direction with decreasing $r$) {}~\footnote{Of course here we refer to ECOs instead of black holes, the label ``hole'' is for matching the notations from~\cite{Teukolsky:1974yv}.} and ``refl'' labeling the right-propagation (reflected) modes (in this paper, ``right'' means toward direction with increasing $r$).

Note that for outgoing modes of either ${}_{+2}\Upsilon$ or ${}_{-2}\Upsilon$, we have
\begin{equation}
e^{-i\omega t}\,e^{i k r_*} \, e^{im\phi} = e^{-i\omega (t-r_*)} \, e^{im(\phi-\Omega_H r_*)}\,,
\end{equation}
therefore the outgoing modes are functions of the retarded time $u=t-r_*$, and the position-dependent angular coordinate $\phi-\Omega_H r_*$.  Similarly for ingoing modes, we have
\begin{equation}
 e^{-i\omega t} \, e^{-i k r_*} \, e^{im\phi} = e^{-i\omega (t+r_*)} \, e^{im(\phi+\Omega_H r_*)}\,,
\end{equation}
indicating that the ingoing modes are functions of the advanced time $v=t+r_*$, and another position-dependent angular coordinate $\phi + \Omega_H r_*$.  We can then write down one schematic expression for either ${}_{+2}\Upsilon$ or ${}_{-2}\Upsilon$, by decomposing both of them into left- and right- propagation components: 
\begin{align}
\label{psi0vac}
{}_{+2}\Upsilon (t, r_*, \theta, \phi)=& {}_{+2}\Upsilon^R (u,\theta,\varphi_-) + \frac{1}{\Delta^{2}} {}_{+2}\Upsilon^L (v,\theta,\varphi_+) \,,\\
\label{psi4vac}
{}_{-2}\Upsilon(t, r_*, \theta, \phi) =& {}_{-2}\Upsilon^R (u,\theta,\varphi_-) + \Delta^{2}{}_{-2}\Upsilon^L (v,\theta,\varphi_+) \,,
\end{align} 
where we have defined
\begin{align}
\varphi_{-} &= \phi-\Omega_H r_* \,,& \varphi_+ &= \phi + \Omega_H r_* \,.
\end{align}
Here both the $L$ and $R$ components are finite, and the $\Delta$ represents the divergence/convergence behaviors of the components.  As we can see here, once we specify these $L$, and $ R$ components on a constant $t$ slice, as functions of $(r_*,\theta,\phi)$, we will be able to obtain their future, or past, values by inserting $t$.  

Here we also note that, while the {\it vacuum/homogeneous} perturbation of space-time geometry are encoded in both $\psi_0$ and $\psi_4$ --- either of them suffices to describe the perturbation field~\cite{Starobinskil:1974nkd,Teukolsky:1974yv}~\footnote{One may imagine a very rough Electromagnetic analogy: for a vacuum EM wave (without electro- or magneto-static fields), both $E$ and $B$ fields contain the full information of the wave, since one can use Maxwell equations to convert one to the other.  Nevertheless, when it comes to interacting with charges and currents, $E$ and $B$ play very different roles, and sometimes it is important to evaluate both $E$ and $B$ fields. }. Near the horizon, the numerical value of $\psi_0$ is dominated by left-propagating waves, while the numerical value of $\psi_4$ is dominated by right-propagating waves. According to Eq.~\eqref{eq:Emm-to-psi}, we then have
\begin{equation}
\mathcal{E}_{mm} \approx -\frac{1}{4\Sigma\Delta} {}_{+2}\Upsilon^L (v,\theta,\varphi_+) - \frac{{\rho^*}^4 \Sigma }{\Delta} \lrsbrk{{{}_{-2}\Upsilon^R} (u,\theta,\varphi_-)}^* \,.
\end{equation}
Note that both terms diverge toward $r_*\rightarrow -\infty$ and at the same order.  This divergence correctly reveals the fact that the FIDOs will observe gravitational waves with the same fractional metric perturbation, but because the frequency of the wave gets increased, the curvature perturbation will diverge as $\alpha^{-2}$.

\subsection{Linear Response Theory}
\label{subsec:linres}

Now, suppose we have a surface, $\mathcal{S}$ at a constant radius $r_* =b_* $ (or in the Boyer-Lindquist coordinates $r = b$), with $e^{\kappa b_*} \ll 1$. Here $\kappa = (r_H -r_C)/2(r^2_H + a^2)$ is the surface gravity of the Kerr black hole. To the right of the surface, for $r_* >  b_* $, we have completely vacuum, and to the left of the surface, we have matter that are relatively at rest in the FIDO frame, we shall refer to this as the ECO region.  The ECO is assumed to be extremely compact and $\mathcal{S}$ is close to the position, as viewed as part of its external Kerr spacetime. 

For the moment, let us  assume that linear perturbation theory holds throughout the external Kerr spacetime of the ECO.  On $\mathcal{S}$ and to its right, $\mathcal{E}_{mm}$ will be the sum of two pieces, 
\begin{equation}
\mathcal{E}_{mm} =\mathcal{E}^{\rm src}_{mm}  + \mathcal{E}^{\rm resp}_{mm}\,,
\end{equation} 
with the first term
\begin{equation}
\mathcal{E}^{\rm src}_{mm}= - \frac{ \Delta }{4\Sigma} {}_{+2}\Upsilon^{\rm src} (v,\theta,\varphi_+)
\end{equation}
a purely left-propagating wave that is {\it sourced} by processes away from the surface, e.g., an orbiting or a plunging  a particle.     The second term can be written as 
\begin{align}
\mathcal{E}^{\rm resp}_{mm} =- \frac{{\rho^*}^4 \Sigma}{\Delta} \left[ {}_{-2}\Upsilon^{\rm refl} (u,\theta,\varphi_-)\right]^*\,,
\end{align} 
as the ECO's {\it response} to the incoming gravitational wave. 

Now we are prepared to discuss the reflecting boundary condition of the Teukolsky equations in terms of the  tidal response of the ECO.  
According to the linear response theory, we can assume the linear tidal response of the ECO is proportional to the total tidal fields near the surface of the ECO.  That is, we may introduce a new parameter $\eta$, and write
\begin{equation}
\mathcal{E}^{\rm resp}_{mm} = \eta(b,\theta) \, \mathcal{E}_{mm}\,.
\end{equation} 
Here $\eta$ is analogous to the {\it tidal love number}.  This leads to the following relation at $r_* =b_*$:
\begin{align}
\label{eq:boundary}
\frac{\left[{}_{-2}\Upsilon^{\rm refl} (t - b_*,\theta,\phi-\Omega_H b_*)\right]^*
}{{}_{+2}\Upsilon^{\rm src} (t  +b_*,\theta,\phi+\Omega_H b_*)}=\frac{\eta}{1-\eta} \frac{e^{-4i \beta}}{4} \Delta^2 \,.
\end{align}
In particular, when $\eta \rightarrow \infty$, we will have the Dirichlet boundary condition, 
\begin{equation}
\label{eq:boundary-d}
\frac{\left[{}_{-2}\Upsilon^{\rm refl} (t - b_*,\theta,\phi-\Omega_H b_*)\right]^*
}{{}_{+2}\Upsilon^{\rm src} (t  +b_*,\theta,\phi + \Omega_H b_*)}=-\frac{e^{-4i \beta}}{4}\Delta^2  \,.
\end{equation}
This then provides us with a prescription for obtaining the boundary condition at $r_*=b_*$.  Once we know the left-propagating $\psi_0^{\rm src}$, the reflected waves due to the tidal response are simply given by Eq.~\eqref{eq:boundary}.

Let us now define a new parameter $\mathcal{R}(b,\theta)$ as
\begin{equation}
\mathcal{R}(b,\theta) \equiv  - \frac{\eta}{1-\eta} \,. 
\end{equation}
This parameter has the physical meaning of being the \textit{ reflectivity} of the tidal fields on the ECO surface.
This {\it local} response assumes that different angular elements of the ECO act independently, which is reasonable since on the ECO surface, and in the FIDO frame, the gravitational wavelength is blue shifted by $\alpha$, hence, much less than the radius of the ECO.  

However, the response may also depend on the history of the exerted tidal perturbation, and  to account for this we can write a more general boundary condition on the ECO surface as
\begin{widetext}
\begin{align}
\label{eq:ref-model}
&\!\!\left[{}_{-2}\Upsilon^{\rm refl} (t - b_*,\theta,\phi-\Omega_H b_*)\right]^* = - \frac{e^{-4i \beta} }{4} \int_{-\infty}^t   dt' \mathcal{R}(b,\theta;t-t') \Delta^2  {}_{+2}\Upsilon^{\rm src} (t'+b_*,\theta,\phi + \Omega_H b_* - \Omega_H(t-t^\prime))\,.
\end{align}
\end{widetext}
Here the $-\Omega_H(t-t')$ term has been inserted into the argument of $_{+2}\Upsilon^{\rm src}$ because the FIDO follows $\phi  =\phi_0 +\Omega_H t$ (see Fig.~\ref{fig:surface}).  This is the key equation of our reflection model.

\subsection{Mode Decomposition}
\label{subsec:mode}

We now have obtained the modified boundary condition~\eqref{eq:ref-model} in terms of the Newman-Penrose quantities, and are ready to apply it to the Teukolsky formalism. The solutions to the $s=-2$ Teukolsky equation, ${}_{-2}\Upsilon$, admits the near-horizon decomposition as in Eq.~\eqref{eq:Teukolsky-H-decompose}.  In this equation, $ Z^{\rm hole} $ is the amplitude of the ingoing wave down to the ECO, which is contributed by the source, and $Z^{\rm refl}$ is the amplitude of the reflected wave due to the tidal response.  For $s=+2$, the corresponding amplitudes are $ Y^{\rm hole} $ and $ Y^{\rm refl} $.  We would like to derive a relation among  the four amplitudes.  

Near the ECO surface $\mathcal{S}$, ${}_{+2}\Upsilon^{\rm src}$ is given by
\begin{align}
{}_{+2}\Upsilon^{\rm src} (v,\theta,\varphi_+)
                          & = \sum_{\lm}  \int \frac{d \omega}{2\pi }e^{-i\omega v}  Y^{\rm hole}_{\lmo} \, \Delta^{-2}
\,{}_{+2}S_{\lmo}(\theta,\varphi_+) \,,
\end{align}
where we have kept only the dominant piece---the left-propagating mode, and $ Y^{\rm hole}_{\lmo}$ is the amplitude of that mode.  The quantity ${}_{-2}\Upsilon^{\rm refl}$ is given by 
\begin{align}
{}_{-2}\Upsilon^{\rm refl} (u,\theta,\varphi_-)
&=\sum_{\lm}  \int \frac{d \omega}{2\pi }e^{-i\omega u}  Z^{\rm refl}_{\lmo}
\,{}_{-2}S_{\lmo}(\theta,\varphi_-) \,,
\end{align}
Inserting the above two equations into Eq.~\eqref{eq:ref-model}, we obtain
\begin{align}
\label{bcfull}
&\sum_{\ell} Z_{\lmo}^{\rm refl} \,{}_{-2}S_{\lmo}(\theta,\phi)   \nonumber\\
= &\frac{1}{4}\sum_{\ell^\prime} e^{4i\beta -2ikb_*} (-1)^{m+1} \mathcal{R}^*_{-k^*} \, Y^{\rm hole^*}_{\lpmmmos} \,{}_{-2}S_{\lpmos}(\theta,\phi)\,,
\end{align}
 where $\mathcal{R}_{\omega}(b,\theta)$ is the Fourier transform of $\mathcal{R}(b,\theta;t-t^\prime)$, and $k\equiv \omega- m \Omega_H$. 
 During the derivation, we have used the fact that the spheroidal harmonic functions satisfy the relation
\begin{equation}
\label{eq:s-relation}
{}_{-2}S^*_{\lmo} (\theta,\phi) = (-1)^m {}_{+2}S_{\lmmmos} (\theta,\phi)\,.
\end{equation}
Assuming the normalization that 
\begin{equation}
\int_0^\pi \vert {}_{-2}S_{\lmo} (\theta) \vert^2 \sin\theta d\theta = 1\,,
\end{equation}
from Eq.~\eqref{bcfull}, we can write 
\begin{equation}
\label{generalbc}
Z_{\lmo}^{\rm refl} =  (-1)^{m+1} \frac{1}{4} e^{-2ikb_*}\sum_{\ell'} \mathcal{M}_{\ell \ell'm\omega}  Y^{\rm hole^*}_{\lpmmmos}  \,,
\end{equation}
with 
\begin{align}
\label{eq:mode-mix-factor}
\mathcal{M}_{\ell \ell'm\omega} &= \int_0^\pi \mathcal{R}^*_{-\omega^* + m \Omega_H}(b,\theta)\,e^{4i\beta(\theta) } \times  \\ \nn
& \times{}_{-2}S_{\lpmo}(\theta) {}_{-2}S^*_{\lmo} (\theta) \sin\theta \, d\theta\,.
\end{align}
In general the reflection will mix between modes with different $\ell$, but not different $m$.  Note that the mixing not only arises from the $\theta$ dependence of $\mathcal{R}(\theta,b)$, but also from the $\theta$ dependence of $\beta$.  This mixing vanishes for the Schwarzschild case.  
For our calculation, it will be good to discard the phase term $e^{4i\beta}$ and make the assumption that $\mathcal{R}$ is independent of the angle $\theta$.  But we should keep in mind that these assumptions only work well in the Schwarzschild limit $a \rightarrow 0$.

In the simplified scenario where mode mixing is ignored, we can write 
\begin{equation}
Z^{\rm refl}_{\lmo} \approx \lrbrk{-1}^{m+1} \frac{1}{4} \, e^{-2 i k b_*}\,  \mathcal{R}^*_{-\omega^* + m \Omega_H} \, Y^{\rm hole^*}_{\lmmmos} \,,
\end{equation}
In this way, the  $\omega$ frequency component of the $\psi_4$ amplitude of each $(l,m)$ mode is related to the $-\omega^*$ frequency component of $\psi_0$ of the $(l,-m)$ mode.   Here in a Fourier analysis, $\omega$ is always real, but we have kept $\omega^*$ so that our notation will directly apply to quasi-normal modes, where frequency can be complex.

\section{Wave Propagation in the vicinity of the horizon}
\label{sec:R-in-emergent-G}

In the previous section, we have obtained a new reflecting boundary condition~\eqref{eq:ref-model} relating Newman-Penrose quantity $\psi_0$ and $\psi_4$ on a spherical surface near the Kerr horizon.  This was further converted as a relation between frequency components of the in-coming $\psi_0$ and the out-going $\psi_4$.  Before moving on to the applications of these boundary conditions, in this section, we put the discussions of the previous section onto a more solid ground.  We consider a concrete coordinate system associated with the FIDOs, and relate condition~\eqref{eq:ref-model} to modified refractive indices or dispersion relations of gravitational waves in this coordinate system. This way of modeling the ECO  can be thought of as a generalization of Refs.~\cite{Wang:2019rcf,Oshita:2018fqu,Oshita:2019sat} to the spinning case.

\subsection{Rindler approximations}

Let us now study the propagation of waves near the horizon, and explore how emergent gravity might influence the boundary condition there.

Inside the ECO boundary $\mathcal{S}$, we can consider propagation of metric perturbations in the near-horizon FIDO coordinate system. According to MP, the unperturbed metric takes the simple form~\cite{Thorne:1986iy}:
\begin{equation}
d s^2 =-\alpha^2 d \tbar^2 +\frac{d\alpha^2}{g_H^2}+\Sigma_{H} d \bar\theta^2 +\frac{4r_H^2}{\Sigma_H} \sin^2 \bar\theta d\bar\phi^2\,,
\label{eq:Rindler-metric}
\end{equation}
where 
\begin{align}
&g_H = \frac{r_H-1}{2 r_H}\,, & \Sigma_H = r^2_ H+ a^2 \cos^2\bar{\theta}\,.
\end{align}
This metric, only valid for $\alpha\ll1$, is a Rindler-like spacetime with spherical symmetry, with horizon located at $\alpha =0$. According to the membrane paradigm~\cite{Suen:1988kq}, the new radial coordinates $(\alpha,\,\bar{\theta},\,\bar{\phi})$ are defined as 
\begin{align}
\tbar & = t \,, \\ 
\alpha &= \lrbrk{2g_H-2a\Omega_H g_H \sin^2\theta}^{\frac{1}{2}} \lrbrk{r-r_H}^{\frac{1}{2}} \,, \\
\bar\theta&=\theta-\frac{{\Sigma_H}_{,\theta}}{4g_H^2 \Sigma_H^2}\alpha^2\,, \\
\bar\phi &=\phi-\Omega_H t \,.
\end{align}
The Kinnersley tetrad, near the horizon, can then be expressed in terms of the Rindler coordinates as
\begin{align}
\vec l &=\frac{2 r_H}{\Delta} (\vec\partial_{\tbar} +g_H\alpha\vec\partial_\alpha) \,, \\
\vec n &=\frac{r_H}{\Sigma} (\vec\partial_{\tbar} -g_H\alpha\vec\partial_\alpha) \,,\\
\vec m & =\frac{-\rho^*}{\sqrt{2}}
\Big[
ia\sin\bar\theta \vec\partial_{\tbar}
-\frac{a\Omega_H\sin\bar\theta\cos\bar\theta}{1-a \Omega_H \sin^2\bar\theta} \alpha\vec\partial_\alpha \nonumber\\
&\qquad\quad +\vec\partial_{\bar\theta} +\left(\frac{i}{\sin\bar\theta}-ia \Omega_H \sin\bar\theta\right)\vec\partial_{\bar\phi}
\Big]\,,
\end{align}
where we have used the near-horizon approximations and discarded all $\mathcal{O}(\alpha^2)$ corrections.

For convenience, we introduce a new radial coordinate $x$, which is related to the lapse function via
\begin{equation}
\label{eq:eqalphax}
\alpha = e^{g_H x} \,.
\end{equation}
The regime $x\rightarrow -\infty$ is the horizon, where $\alpha\rightarrow 0$.  In fact, $(t,x)$ is exactly the Cartesian coordinate of the Minkowski space in which this Rindler space is embedded.
Now we consider metric perturbations of the trace-free form
\begin{align}
h_{\bar\theta\bar\theta}(t,x,\bar\theta,\bar\phi)\ &= \Sigma_H  H_+(t,x,\bar\theta,\bar\phi)\,,\\
h_{\bar\theta\bar\phi}(t,x,\bar\theta,\bar\phi)&= 2 r_H\sin\bar\theta  H_\times (t,x,\bar\theta,\bar\phi)\,,\\ 
h_{\bar\phi\bar\phi}(t,x,\bar\theta,\bar\phi) &=-4 r^2_H\sin\bar\theta^2 H_+(t,x,\bar\theta,\bar\phi) /\Sigma_H  \,.
\end{align}
Note that $H_{+,\times}$ are metric perturbations in the angular directions, measured in orthonormal bases.  We first find that the Einstein's equations reduce to 
\begin{equation}
\label{eq:LEE}
(-\partial_t^2+\partial_x^2 ) H_p =0\,,\quad p=+,\times.
\end{equation}
Again, to obtain the equations above we have only kept the leading terms in $\alpha$-series.  The absence of $\bar\theta$ and $\bar\phi$ derivatives in this equation supports the argument that tidal response of the ECO is {\it local} to each angular element on its surface, as we have made in Sec.~\ref{subsec:linres}.

We can further decompose $H_p(t,x)$ to the left- and right-propagating piece as 
\begin{equation}
H_p(t,x,\bar\theta,\bar\phi) = H_p^L(t+x,\bar\theta,\bar\phi) +H_p^R(t-x,\bar\theta,\bar\phi) \,,
\end{equation}
Using the Rindler approximations, we then find that the Weyl quantities $\psi_0$ and $\psi_4$ near horizon can be written as 
\begin{align}
\label{eq:psi0-rindler}
\psi_0(t,x,\bar\theta,\bar\phi)& = \frac{8r_H^2 e^{2i\beta}}{\Delta^2}(\partial_t^2-g_H\partial_x)\mathcal{H}^{L}(t,x,\bar\theta,\bar\phi) \,,\\
\label{eq:psi4-rindler}
\Big[\rho^{-4} \psi_4 (t,x,\bar\theta,\bar\phi)\Big]^*  &= 2r_H^2 e^{-2i\beta}(\partial_t^2-g_H\partial_x)\mathcal{H}^{R}(t,x,\bar\theta,\bar\phi)\,,
\end{align}
where we have defined 
\begin{align}
\mathcal{H}^{L} &= H^{L}_++i H^L_\times \,, &
\mathcal{H}^{R} &= H^{R}_++i H^R_\times \,.
\end{align}
Note that in this approximation, we only extract the leading behavior of $\psi_0$ and $\psi_4$ near the horizon, namely a left-going wave $\sim (r-r_H)^{-2}$ for $\psi_0$, and a right-going wave $\sim (r-r_H)^0$ for $\psi_4$.  Here, we are considering wave propagation and reflection independently for each $(\bar\theta,\bar\phi)$. Eq.~\eqref{eq:psi0-rindler} and Eq.~\eqref{eq:psi4-rindler} are consistent with our reflection model given in Eq.~\eqref{eq:ref-model}.  For instance, in the case of total reflection, we have $\mathcal{R}=-1$, and all left-propagating modes $H^{L}_p$ become right-propagating modes $H^{R}_p$.

\begin{figure}[t]
\includegraphics[width=0.4\textwidth]{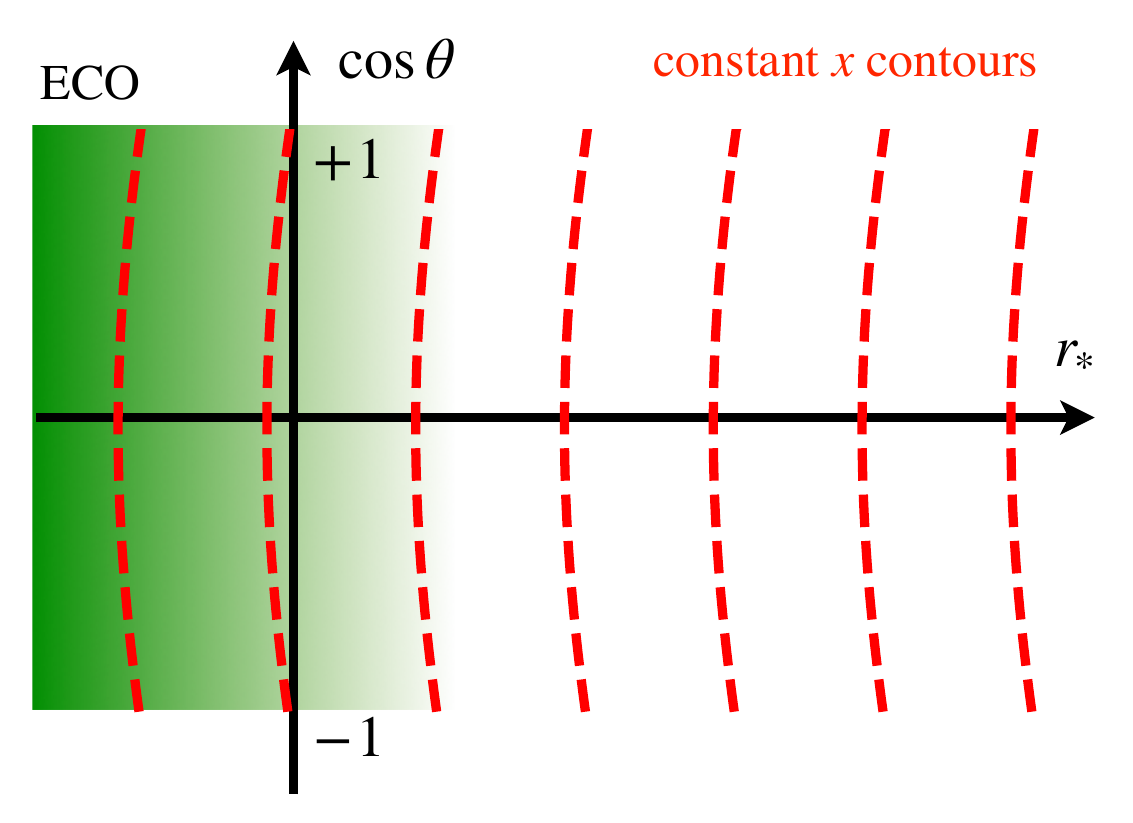}
\caption{Illustration of the constant-$x$ contours in the $(r_*,\cos\theta)$ plane.   Reflections from the same $x$ for different $\theta$ will appear as being reflected at different $r_*$ for different $\theta$. \label{fig:xcontours}}
\end{figure}

Let us now evaluate the Riemann tensor components in an orthonormal basis whose vectors point along the $(t,x,\bar\theta,\bar\phi)$ coordinate axes. The results are  
\begin{align}
R_{\hat{\tbar} \hat{\bar\theta} \hat{\tbar} \hat{\bar \theta}}=-R_{ \hat{\tbar} \hat{\bar \phi} \hat{\tbar} \hat{ \bar \phi}}  &= -\frac{e^{-2g_H x}}{2}\left[-\partial_t^2+g_H\partial_x\right] H_+ \,, \\
R_{\hat{\tbar} \hat{\bar \theta} \hat{\tbar} \hat{ \bar \phi}} &= -\frac{e^{-2g_H x}}{2}\left[-\partial_t^2+g_H\partial_x\right] H_\times\,.
\end{align}
This also confirms the reflection model that we have obtained from the previous section.

We also point out that near the horizon, $x$ and $r_*$ differ by a additive constant for each $(\bar\theta,\bar\phi)$.  Let's work out the $\bar \theta$ dependence of the  asymptotic shift between $x $ and $r_*$.
More specifically, near the horizon, the tortoise coordinate $r_*$ is approximately given by
\begin{equation}
r_* \approx  \frac{1}{2g_H} \ln [2g_H(r-r_H)] + \mathcal{I}\,,
\end{equation} 
with a constant
\begin{equation}
\mathcal{I} = r_H + \ln\lrbrk{\frac{r_H}{2}}  - \frac{1}{2r_H g_H}\ln \lrbrk{8 r_H g^2_H} \,.
\end{equation}
Here we have neglected $\mathcal{O}(r-r_H)$ terms.
Note that $r_*$ is independent of $\bar\theta$. We define the difference between the two radial coordinates as 
\begin{equation}
 x -r_* \equiv \delta (\bar\theta) - \mathcal{I} \,,
\end{equation}
where
\begin{equation}
\delta (\bar\theta) = \frac{1}{2g_H} \ln(1-a\Omega_H\sin^2\bar\theta)\,.
\end{equation}
This may influence the mode mixing of reflected waves from an ECO whose surface has a constant redshift.   In Fig.~\ref{fig:xcontours}, we illustrate contant-$x$ contours in the $(r_*,\cos\theta)$ plane.

Finally let us derive the Teukolsky reflectivity $\mathcal{R}$ using the Rindler approximation. Supposing for $x$ in certain regions we can write the wave solution as 
\begin{align}
\label{eq:H-reflect}
\mathcal{H}(t,x,\bar\theta, \bar\phi)= \mathcal{H}^L(t,x,\bar\theta, \bar\phi) + \mathcal{H}^R(t,x,\bar\theta, \bar\phi) \,,
\end{align}
with
\begin{align}
\mathcal{H}^L(t,x,\bar\theta, \bar\phi) & = \sum_m \int \frac{d k}{2\pi} \Theta_k(\bar\theta)  e^{-i k x} e^{-i k t}e^{im\bar\phi} \,, \\ 
\mathcal{H}^R(t,x,\bar\theta, \bar\phi) & = \sum_m \int \frac{d k}{2\pi} \Theta_k(\bar\theta)  \zeta_k e^{i k x} e^{-i k t}e^{im\bar\phi}\,.
\end{align}
Here $\Theta_k(\bar\theta)$ gives the $k$-dependent angular distribution. $\zeta_k \equiv \zeta(k)$ is the reflection coefficient that converts left-propagating to right-propagating gravitational waves.  Thus $\psi_0$ and $\psi_4$ are respectively given by
\begin{widetext}
\begin{align}
\psi&_0 (t, x, \bar\theta, \bar\phi) = \frac{8 r^2_H e^{2i\beta(\bar\theta)}}{\Delta^2}  \sum_m \int \frac{d k}{2\pi} \Theta_k(\bar\theta) (-k^2 +i g_H k)e^{-i k x} e^{-i k t}e^{im\bar\phi} \,, \\
(\rho^{-4}\psi&_4)^* (t, x, \bar\theta, \bar\phi) = 2r_H^2 e^{-2i\beta(\bar\theta)} \sum_m \int \frac{d k}{2\pi} \Theta_k(\bar\theta) (-k^2 -i g_H k)  \zeta_k e^{i k x} e^{-i k t}e^{im\bar\phi} \,.
\end{align}

Now that we have obtained $\psi_0$ and $\psi_4$ using the Rindler approximations.  We would like to relate $\zeta$ to the Teukolsky reflectivity $\mathcal{R}$. To accomplish this, recall that in Sec.~\ref{sec:near-horizon-BC} we have obtained a reflection relation~\eqref{eq:ref-model} between $\psi_0$ and $\psi_4$ on the ECO surface. Since the relation is written in the Boyer-Lindquist coordinates, we first perform the coordinate transformations on $\psi_0$ and $\psi_4$ according to $x = r_* +\delta(\theta) - \mathcal{I}$, $\bar\theta = \theta$, and $\bar \phi = \phi -\Omega_H t$. During the coordinate transformation, we have used the near-horizon approximations and discarded all $\mathcal{O}(\alpha^2)$ terms.  The results are given by
\begin{align}
\psi&_0 (v, \theta,\varphi_+) = \frac{8 r^2_H e^{2i\beta(\theta)}}{\Delta^2}  \sum_m \int \frac{d k}{2\pi} \Theta_k(\theta) (-k^2 +i g_H k)e^{-i (k+m\Omega_H) v} e^{-i k \delta(\theta)} e^{ik\mathcal{I}}e^{im \varphi_+} \,, \\
(\rho^{-4}\psi&_4)^* (u, \theta, \varphi_-) = 2r_H^2 e^{-2i\beta(\theta)}  \sum_m \int \frac{d k}{2\pi} \Theta_k(\theta) (-k^2 -i g_H k)  \zeta_k e^{-i (k + m\Omega_H) u} e^{i k \delta(\theta)} e^{-ik\mathcal{I}} e^{im\varphi_-} \,.
\end{align}

\end{widetext}
Using the reflection model~\eqref{eq:ref-model}, we obtain that 
\begin{align}
\label{eq:zeta-to-R}
\mathcal{R}_k = \zeta_k \lrbrk{\frac{-k -i g_H }{-k + i g_H }} \exp\lrsbrk{2ik b_* + 2ik \delta(\theta) - 2ik\mathcal{I}}\,.
\end{align}
Thus once we know $\zeta$, the Teukolsky reflectivity $\mathcal{R}$ can be readily obtained.  Here we point out that the phase factor $e^{2ikb_*}$ here will cancel the $e^{-2ikb_*}$ factors in Sec.~\ref{subsec:mode}.  This is because in the previous section we chose $b_*$ as  the location for the  ``surface of the ECO'', while in this section, the ECO is embedded into the $x$ coordinate system, therefore we no longer need to introduce a reference location $b_*$ as ``surface of the ECO''.   The information of ECO location will now be incorporated into $\zeta_k$.

\begin{figure*}[htbp!]
\centering
\includegraphics[width=0.48\textwidth]{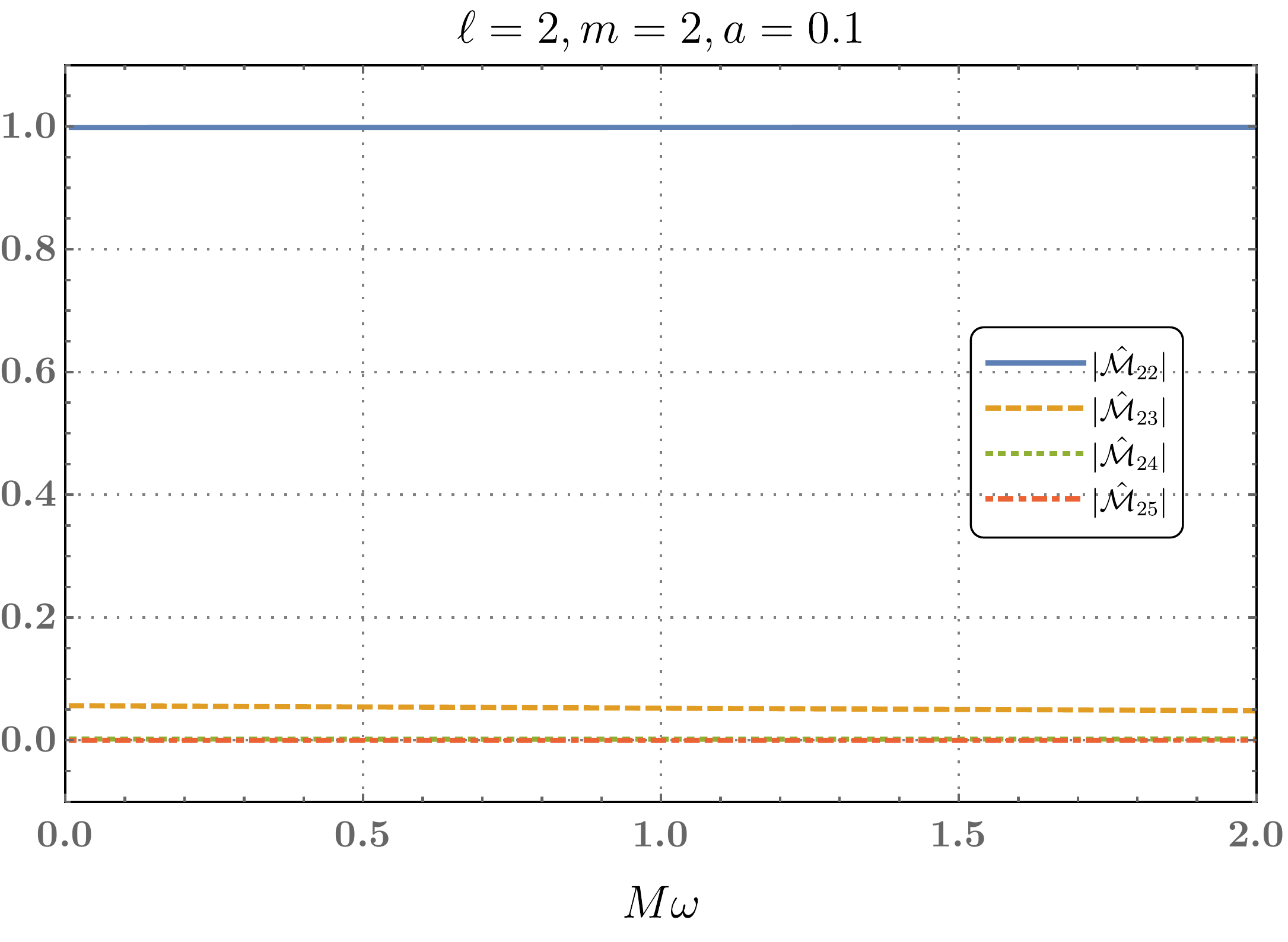}
\includegraphics[width=0.48\textwidth]{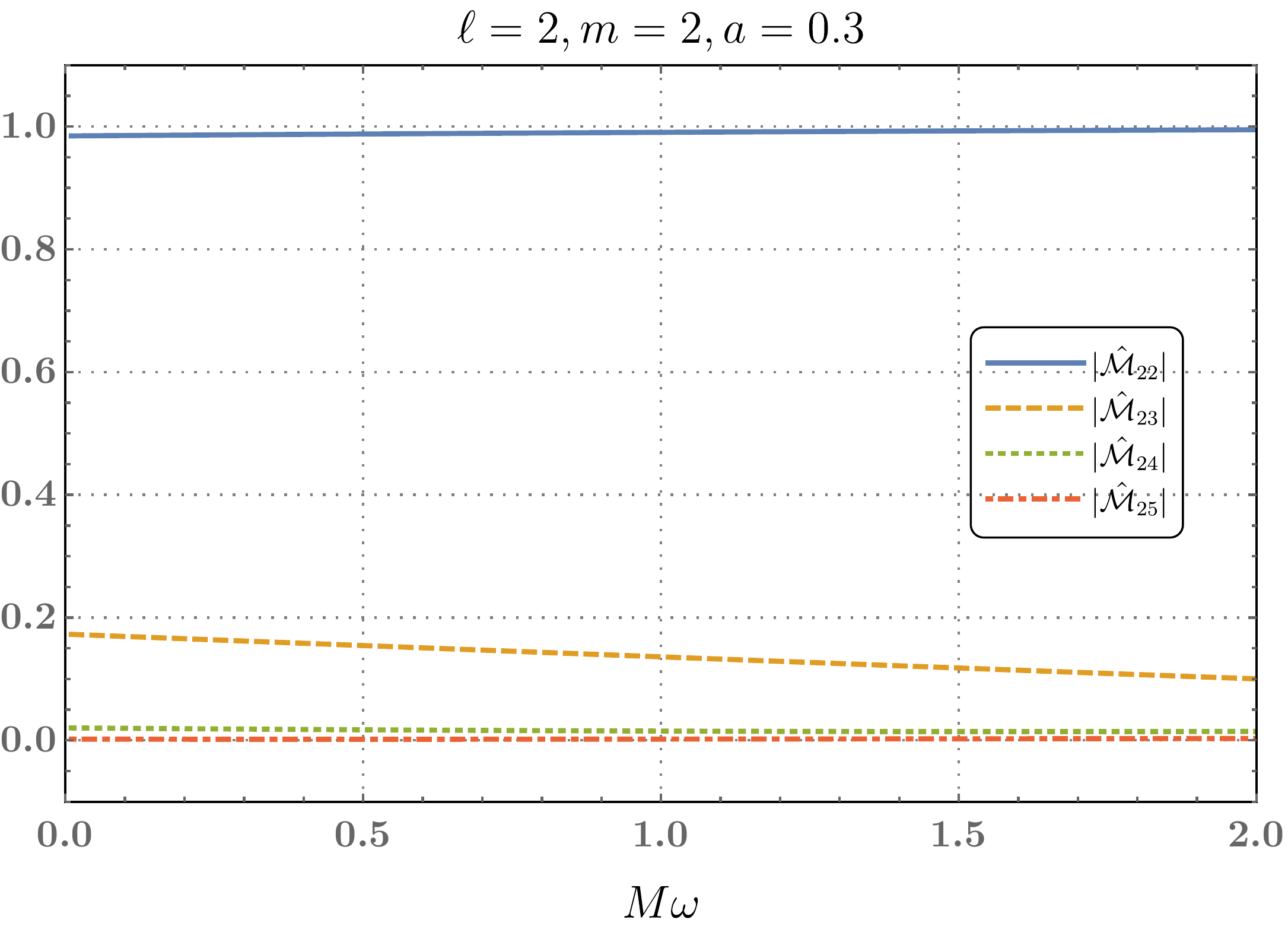}
\includegraphics[width=0.48\textwidth]{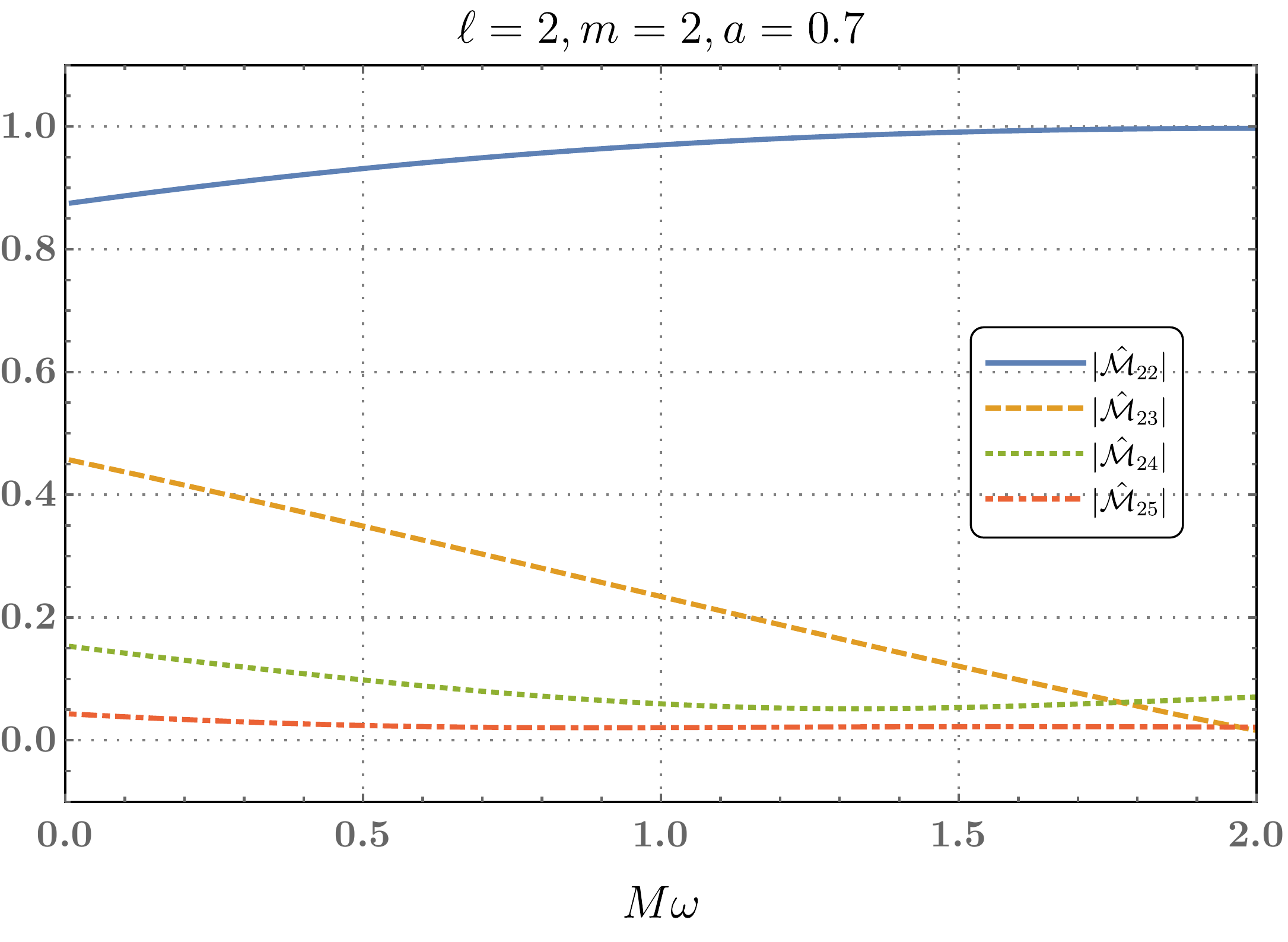}
\includegraphics[width=0.48\textwidth]{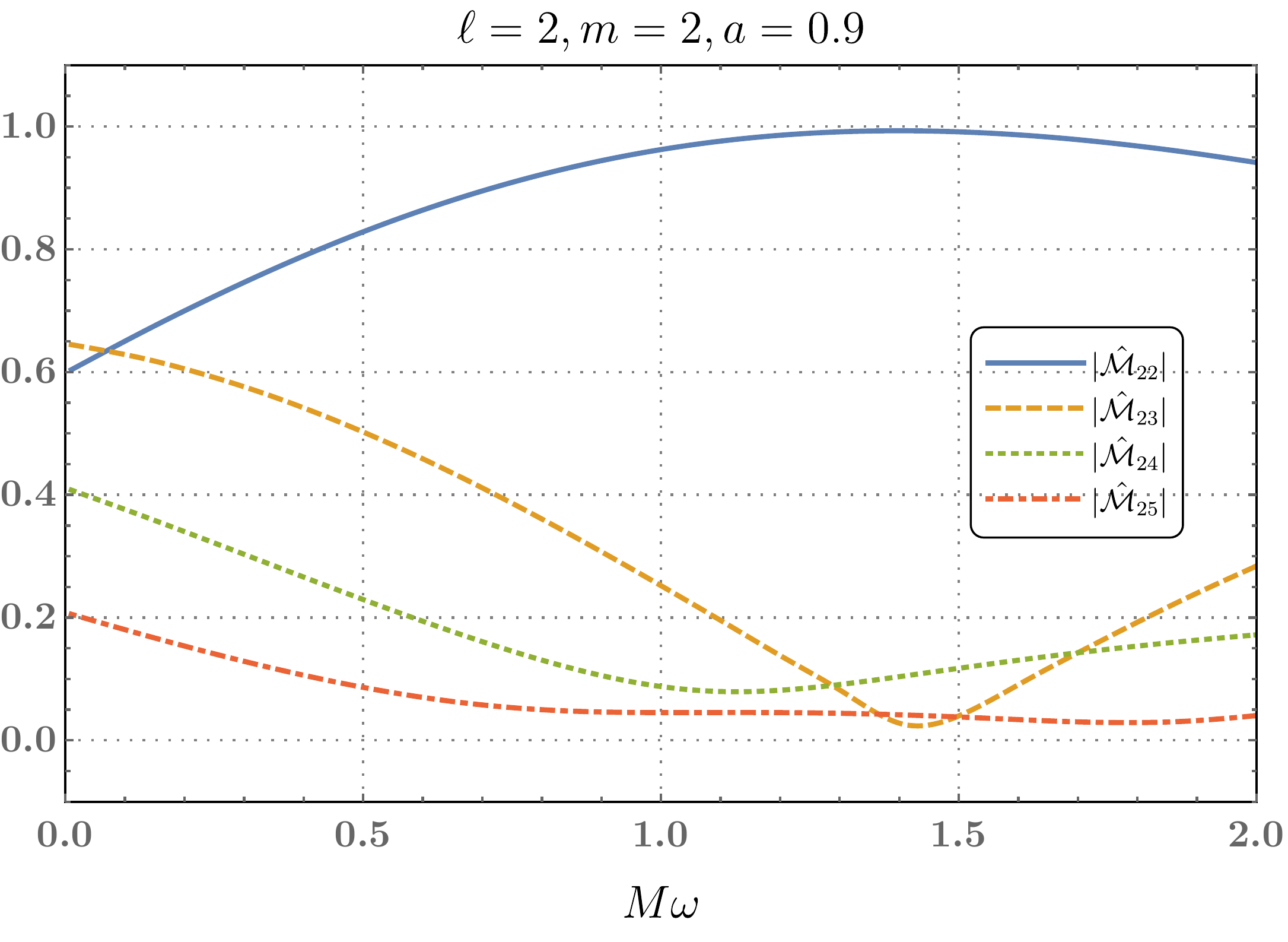}
\caption{ The absolute values of the factor $\hat{\mathcal{M}}_{\ell \ell^\prime m\omega}$ for various spin $a$ and $\ell^\prime$. This factor shows the mixing between different $\ell$-modes after an incoming single mode gets reflected on the surface of an exotic compact object.
Here we have chosen $\ell=2$, $m=2$ as an example.  In general for higher spin the reflected waves gain more contributions from higher $\ell^\prime$-modes, thus the effect of mode mixing are not negligible for rapidly spinning ECOs.}
\label{fig:mode-mix-factor}
\end{figure*}

Before the end of this subsection, let us look at the factor $\mathcal{M}_{\ell \ell'm\omega}$ in Eq.~\eqref{eq:mode-mix-factor}, and see how the mode mixing show up in the reflected waves.  We can pull out the angular dependence of this factor by defining 
\begin{align}
\mathcal{M}_{\ell \ell'm\omega} = \lrbrk{\frac{-k -i g_H }{-k + i g_H }} e^{2ik b_*-2ik\mathcal{I}}\hat{\mathcal{M}}_{\ell \ell'm\omega} \,,
\end{align}
where
\begin{align}
\hat{\mathcal{M}}_{\ell \ell'm\omega} = \int_0^\pi \zeta^*_{-\omega^*+m\Omega_H} e^{i \Phi_{m\omega}(\theta)} 
{}_{-2}S_{\lpmo}(\theta) {}_{-2}S^*_{\lmo} (\theta) \sin\theta \, d\theta\,,
\end{align}
with
\begin{equation}
\Phi_{m\omega}(\theta) = 2(\omega-m\Omega_H)\delta(\theta) + 4\beta(\theta)\,.
\end{equation}
As an example,  we simply let $\zeta=1$, thus $\hat{\mathcal{M}}_{\ell \ell'm\omega}$ directly shows the mixing of modes  due to the phase.   We plot the absolute value of $\hat{\mathcal{M}}_{\ell \ell'm\omega}$ for $\ell=2$, $m=2$ and for various spin and $\ell^\prime$ in Fig.~\ref{fig:mode-mix-factor}. For $a=0$, we have $\hat{\mathcal{M}}_{\ell \ell'm\omega}=1$, indicating no mode mixing.  As we raise the spin, modes get more mixed and the reflected waves attain more contributions from $\ell^\prime >2$ modes.  This quantitatively shows the mixing of different $\ell$-modes is a significant feature for reflection of waves near horizon.

\subsection{Position-dependent damping of gravitational waves}

We now calculate the reflectivity $\mathcal{R}$ in a simplest setting --- by adding dissipating terms in the linearized Einstein equation in the Rindler coordinate system, obtaining $\zeta$, and then converting to $\mathcal{R}$. 
Wang et al.\ already introduced a model in which wave is damped, by introducing a complex ``Young's modulus'' of space-time~\cite{Wang:2019rcf}.  They name the reflection coefficient they found as the \textit{Boltzman reflectivity}. As an alternative approach, let us introduce a position-dependent damping to gravitational waves, which increases as we approach the horizon.  This model has the feature of being able to provide more well-posed differential equations. 

To do so, we modify the linearized Einstein equation by adding an extra dissipation term, with coupling coefficient $\epsilon$, to the equation satisfied by the perturbation $\mathcal{H}$ defined in Eq.~\eqref{eq:H-reflect}:
\begin{equation}
\label{eq:damping-eom}
-\partial_t^2  \mathcal{H}-\epsilon e^{-g_H x} \partial_t    \mathcal{H} + \partial_x^2    \mathcal{H}=0\,.
\end{equation}
Assuming harmonic time decomposition $\mathcal{H}(x,t) = \tilde{\mathcal{H}}(x)e^{-ik t}$, we have 
\begin{equation}
\lrsbrk{\frac{d^2 }{dx^2} + k^2 +ik \epsilon e^{-g_H x}} \tilde{\mathcal{H}}(x) = 0\,.
\end{equation}
Here $k$ has the physical meaning of being the angular  frequency of the perturbation measured by FIDOs, before blue shift.  The modified Einstein equation then admits a general solution given by
\begin{equation}
\tilde{\mathcal{H}}(x) = C_1 \Gamma(1-i\nu) J^{(1)}_{-i\nu}(z) + C_2 \Gamma(1+i\nu) J^{(1)}_{i\nu}(z) \,,
\end{equation}
where 
\begin{align}
\nu &= 2k/g_H \,, &
z &= 2 e^{\frac{i\pi}{4}-\frac{g_H x}{2}} \sqrt{\epsilon k}/g_H \,,
\end{align}
and $J^{(1)}_{\nu}(z)$ is the Bessel function of the first kind.
The appropriate solution which damps on the horizon is given by 
\begin{equation}
\frac{C_1}{C_2} = - \frac{\Gamma(1+i\nu)}{\Gamma(1-i\nu)} e^{-\pi \nu}\,.
\end{equation}
Here we shall assume $\epsilon \ll 1$.  In this way, there is a region where $x \ll -1$, but still with $\epsilon e^{-g_H x} \ll 1$. In other words, this is a region very close to the Kerr horizon, but here the damping has not yet turned on.  In this region,   the damping solution can be written as
\begin{equation}
\label{eq:h-decay-decompose}
\tilde{\mathcal{H}}(x)\propto e^{-ik x} + \zeta_{\rm D} e^{ik x}\,,
\end{equation} 
with 
\begin{align}
\label{eq:R-with-dissipation}
\zeta_D (k)= -\frac{\Gamma(1+2ik/g_H)}{\Gamma(1-2ik/g_H)} e^{-\frac{2i k}{g_H} \ln\frac{k}{g_H}} e^{-\frac{\pi k}{g_H}}  e^{-\frac{2ik}{g_H} \ln\epsilon}  \,,\; k>0\,.
\end{align}
Here we have imposed $k>0$ because the sense of in-going and out-going waves changes for $k<0$, where we need to write
\begin{equation}
\zeta_D(-k) =\zeta_D^*(k)\,.
\end{equation}
This is the same form of reflectivity proposed by Wang {\it et al.} In Eq.~\eqref{eq:R-with-dissipation}, the first factor involving two $\Gamma$ functions is a pure phase factor that has a moderate variation at the scale $k\sim g_H$, and  the phase factor $e^{-\frac{2i k}{g_H} \ln\frac{k}{g_H}} $ is similar; the amplitude factor $e^{-\frac{\pi k}{g_H}}$ provides unity reflectivity for $k\sim 0$ and this reflectivity decreases as $|k|$ increases.   We plot $\vert \zeta_D (k) \vert$ for $g_H=1$ in Fig.~\ref{fig:zeta}.

The final phase factor in $\zeta_D$ can be written in the form of
\begin{equation}
e^{-\frac{2ik}{g_H} \ln\epsilon}  =e^{ -2ik x_{\rm eff}}\,,\quad 
 x_{\rm eff} = \frac{1}{g_H} \ln \epsilon\,.
 \end{equation} 
 This provides an effective $x$ location around which most of the wave is reflected --- as we can see, we no longer have a single location $r= b$ for the ECO surface at which all the waves are reflected.    To obtain the reflectivity $\mathcal{R}$, we simply insert $\zeta_D$ into Eq.~\eqref{eq:zeta-to-R}, which adds an additional $\theta$-dependent phase factor.

\subsection{GW Propagation in Matter}

The damping term in the linearized Einstein equation causes reflection in the near-horizon region.  In this subsection we consider another scenario where there exists some effective matter fields in the vicinity of the horizon. The effective stress-energy tensor is denoted as $T_{AB}^{\rm eff}$, and its existence may be related to the emergent nature of gravity.
We now modify the linearized $(1+1)$-Einstein equation~\eqref{eq:LEE} by adding the effective source, and get
\begin{equation}
-\partial_t^2 \mathcal{H}  + \partial_x^2 \mathcal{H}  =-16 \pi e^{2g_H x} T_{AB}^{\rm eff} \,.
\end{equation}
In this equation, on the left hand side, we have a freely propagating GW in $(1+1)$-Minkowski spacetime, while on the right hand side, we have the effect of emergent gravity. 

\subsubsection{Tidal response of matter}

We now discuss how $T_{\rm AB}^{\rm eff}$ should respond to $\mathcal{H}$.  Suppose theses effective degrees of freedom act as matters that stay at rest in the FIDO frame.  The $AB$ component of the Riemann tensor is given by 
\begin{equation}
R_{tAtB} = \frac{1}{2}\left(-\partial_t^2  +g_H \partial_x\right) \mathcal{H} \,.
\end{equation}
We postulate that the response of the effective matter is given by 
\begin{equation}
T^{\rm eff}_{AB} =\frac{\mu}{8\pi} R_{\tau A \tau B} \,,
\end{equation} 
where $\tau$ is the proper time for the Rindler metric~\eqref{eq:Rindler-metric}, and $\mu$ is a physical coupling constant measured in the local Lorentz frame of the FIDO, which can be dependent on the driving frequency felt by the FIDO.  Thus we have 
\begin{equation}
T^{\rm eff}_{AB} = \frac{\mu}{8\pi \alpha^2 } R_{tAtB} =\frac{\mu}{16\pi \alpha^2}\left(-\partial_t^2 + g_H \partial_x\right) \mathcal{H} \,.
\end{equation} 
Note that the Einstein's equation is now modified into
\begin{equation}
G_{AB} = \mu R_{\tau A \tau B} \,.
\end{equation} 
With the effective stress-energy tensor, the metric equation of motion can now be written as
\begin{equation}
\label{eq:gw-in-matter-eq}
\left[-(1 + \mu) \partial_t^2  +  g_H \mu\partial_x   + \partial_x^2\right] \mathcal{H} =0\,.
\end{equation}
Here $(1+\mu)$ acts as the permeability of gravitational waves in matter, and decreases the speed of gravitational waves.  Now let us consider two kinds of matter distributions for the exotic compact object.

\subsubsection{Homogeneous star}

For a simplest model, let us look at a homogeneous star with uniform $\mu$ in the interior region.
For a frequency independent $\mu$, we can write $ \mathcal{H} \propto e^{-ik t+ i \tilde{k} x}$, and the modified dispersion relation is given by $\tilde{k} = \tilde{k}_+$ or $\tilde{k}_-$, with 
\begin{equation}
\tilde{k}_\pm = \frac{ i g_H \mu }{2}\pm
\sqrt{(1+\mu)k^2- \frac{g_H^2\mu^2}{4}}\,.
\end{equation}
We immediately note that gravitational waves become evanescent when 
\begin{equation}
\vert k \vert \leq \vert k_{\rm th} \vert = \frac{\vert g_H \mu \vert}{2\sqrt{1+\mu}}\,.
\end{equation}
That is, we have a total reflection of all waves below $\omega_{\rm th}$.  Substantial reflection also takes place near the $\omega_{\rm th}$ frequency.  For $\vert k \vert > k_{\rm th} $ and positive $\mu$,  waves will be amplified when propagating towards the $x\rightarrow -\infty$ direction, i.e., towards the horizon.

\begin{figure}[htbp!]
\centering
\includegraphics[width=0.48\textwidth]{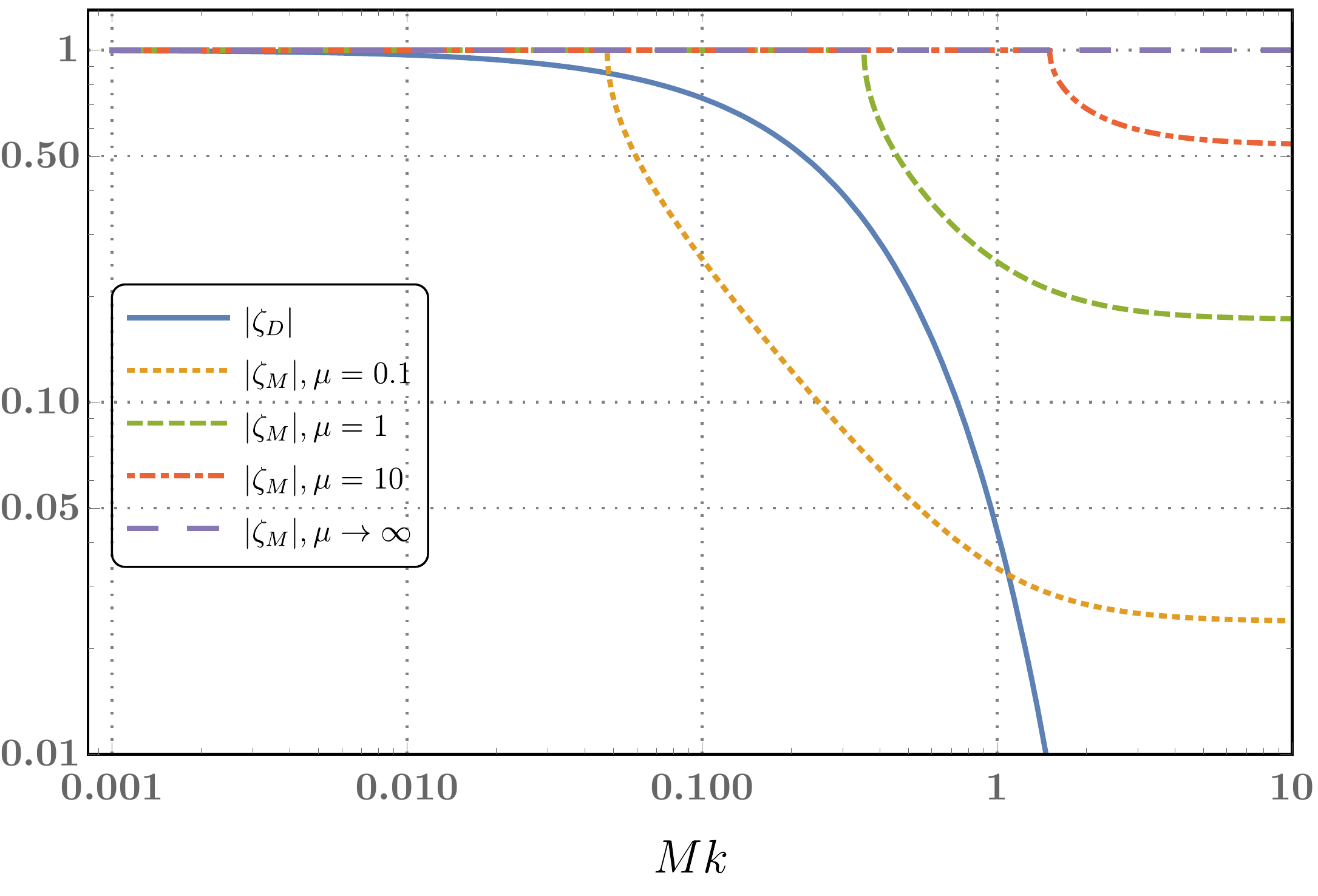}
\caption{Absolute values of $\zeta_D$ and $\zeta_M$ as functions of the frequency $k$. We have set $g_H=1$.  Since $\zeta$ and the Teukolsky reflectivity $\mathcal{R}$ only differ by a phase, $\vert\zeta\vert$ is the same as $\vert \mathcal{R} \vert$.  The blue solid line represents $\vert \zeta_D \vert$. The yellow dotted line,  the green dashed line, the red dot-dashed line, and the purple long dashed line give $\vert \zeta_M \vert$ for $\mu=0.1, 0.2, 0.5, \infty$ respectively.  The ``Boltzman'' reflectivity, i.e. $\xi_D$, exponentially decays for higher frequencies.  For our model of homogeneous stars, we have total reflection of waves on the ECO surface below a certain threshold frequency. Beyond the threshold frequency, the reflectivity gets decreased and converges to a constant.  When $\mu\rightarrow \infty$, we have total reflection of waves for all the frequency range, which is equivalent to the case of inhomogeneous stars we have introduced.   }
\label{fig:zeta}
\end{figure}

We may further postulate that $\mu$ is of order unity inside a surface  at which the surface gravity is blue-shifted to the Planck frequency $\omega_P$:
\begin{equation}
\mu =\left\{
\begin{array}{ll}
\mu_0\,, & \alpha^{-1} g_H > \omega_P\,,\\
\\
0\,, & \mbox{otherwise} \,.
\end{array}\right.
\end{equation} 
The surface is then located at $x=x_P$, where
\begin{equation}
x_P = \frac{1}{g_H}\ln\left(\frac{g_H}{ \omega_P}\right) \,.
\end{equation}
As before, we write down the general solutions to Eq.~\eqref{eq:gw-in-matter-eq} as $\mathcal{H}(x,t) = \tilde{\mathcal{H}}(x)e^{-ik t}$.  
Outside the surface, we can write 
\begin{equation}
\tilde{\mathcal{H}}(x)\propto e^{-ik x} + \zeta_M e^{ik x}\,.
\end{equation} 
Inside the surface we have
\begin{equation}
\tilde{\mathcal{H}}(x)\propto e^{i\tilde{k}_- x}\,.
\end{equation}
Matching the solutions on the surface gives
\begin{equation}
\zeta_M = \lrbrk{ \frac{k  + \frac{ i g_H \mu }{2}-
\sqrt{(1+\mu)k^2- \frac{g_H^2\mu^2}{4}}}{k - \frac{ i g_H \mu }{2} +
\sqrt{(1+\mu)k^2- \frac{g_H^2\mu^2}{4}} } } e^{-2ik x_P} \,,\quad k>0\,.
\end{equation}
Similarly to the previous section, $\zeta_M (-k) = \zeta^*(k)$. For $|k| \le k_{\rm th}$, we have $\vert \zeta_M \vert =1 $, indicating a total reflection of low frequency waves.  For higher frequencies, $\vert \zeta_M \vert $ approaches a constant 
\begin{equation}
\lim_{k\rightarrow \infty}\vert \zeta_M \vert =
\frac{\sqrt{1 + \mu}-1}{1+\sqrt{1 + \mu}} \,.
\end{equation} 
We plot $\vert\zeta_M \vert$ for different $\mu$s in Fig.~\ref{fig:zeta}.
Since $\mu$ is supposed to be a small number, high frequency waves have nearly zero reflection near the surface. This $\zeta_M$ is qualitatively similar to the Lorentzian reflectivity model adopted, e.g., by Ref.~\cite{Du:2018cmp}.

\subsubsection{Inhomogeneous star}

Let us make $\mu$ grow as a function of the location, with
\begin{equation}
\label{eq:mu-model-x-dependent}
\mu = \mu_0 e^{-\eta x} \,,
\end{equation}
where $\mu_0$ and $\eta$ are  positive constants.  In this way, we successfully ``revive'' $\mu$ near the horizon.

We write down the general solutions to Eq.~\eqref{eq:gw-in-matter-eq} as $\mathcal{H}(x,t) = \tilde{\mathcal{H}}(x)e^{-ik t}$, and obtain that
\begin{align}
\tilde{\mathcal{H}}(x) = &A_1 e^{-ik x} e^{\frac{ik}{\eta} \ln\lrbrk{\frac{\mu_0 g_H}{\eta}} - \frac{\pi k}{\eta}} M(a,b,z) + \\ \nn
&A_2 e^{ik x} e^{-\frac{ik}{\eta} \ln\lrbrk{\frac{\mu_0 g_H}{\eta}} + \frac{\pi k}{\eta}} M(a^*,b^*,z)\,,
\end{align}
where 
\begin{align}
a& = \frac{ik}{\eta} - \frac{k^2}{g_H\eta}\,, & b &= 1+ \frac{2ik}{\eta}\,, & z = \frac{\mu_0 g_H}{\eta} e^{-\eta x} \,,
\end{align}
$A_1$, $A_2$ are some constants, and $M(a,b,z)$ is the confluent hypergeometric function.

The hypergeometric function behaves asymptotically as
\begin{align}
M(a,b,z) &\sim e^z \,z^{a-b}\frac{ \Gamma(b)}{\Gamma(a)} \,, & z &\rightarrow \infty\,, \\ 
M(a,b,z) &\sim 1 \,, & z &\rightarrow 0\,.  
\end{align}
The solution that damps on the horizon is then given by
\begin{equation}
\frac{A_2}{A_1} = - e^{-\frac{2\pi\omega}{\eta}}\frac{\Gamma(b)}{\Gamma(b^*)}\frac{\Gamma(a^*)}{\Gamma(a)} \,.
\end{equation}
For $x$ in the region that $\mu_0 g_H e^{-\eta x} \ll 1$ and positive $k$, this solution can then be written as
\begin{equation}
\tilde{\mathcal{H}}(x) = e^{-ik x} + \zeta_N e^{ik x}\,,
\end{equation}
where
\begin{align}
\zeta_N = & - e^{\left[-\frac{2 ik}{\eta }\ln\left(\frac{\mu_0 g_H}{\eta}\right) \right]} 
\frac{\Gamma\left(1 + \frac{2ik}{\eta}\right)}{\Gamma\left(1 - \frac{2ik}{\eta}\right)}
\frac{\Gamma(-\frac{ik}{\eta} -\frac{k^2}{g_H\eta})}{\Gamma(\frac{ik}{\eta}-\frac{k^2}{g_H\eta})}\,.
\end{align}
One immediately notes that $\vert \zeta_N \vert =1$ for all real $k$, indicating a total reflection of waves.  This may due to the fact that our assumption of $\mu$ in ~\eqref{eq:mu-model-x-dependent} is equivalent to putting infinite numbers of reflecting surfaces near the horizon, i.e. the $\mu\rightarrow \infty$ case in Fig.~\ref{fig:zeta}.

\section{Boundary Condition in terms of Various Functions}
\label{sec:boundary-cond-func}

In calculations for gravitational waveforms, one does not usually compute both $\psi_0$ and $\psi_4$; the Sasaki-Nakamura formalism was also used to obtain faster numerical convergence.  In this section, let us convert our boundary condition~\eqref{generalbc}, which  involves both $\psi_0$ and $\psi_4$ amplitudes, into those that only involve $\psi_4$ amplitudes, and compare our reflectivity with the one defined using the Sasaki-Nakamura functions.  

\subsection{Reflectivity for $\psi_4$ mode amplitudes}

The Newman-Penrose quantities $\psi_0$ and $\psi_4$ can be tranformed into each other using the Teukolsky-Starobinsky identities.  The amplitude $Z^{\rm hole}$ and $Y^{\rm hole}$ are related by~\cite{Teukolsky:1974yv}
\begin{equation}
\label{eq:Y-Z-relation}
C_{\lmo} Y_{\lmo}^{\rm hole} = D_{\lmo} Z^{\rm hole}_{\lmo}\,,
\end{equation}
with 
\begin{equation}
D_{\lmo} = 64 (2 r_H)^4 (ik) \lrbrk{k^2 + 4\varepsilon^2} \lrbrk{- ik+ 4\varepsilon}  \,,
\end{equation}
and $C$ is given by
\begin{align}
 \vert C_{\lmo}&\vert ^2
 = \lrbrk{(\lambda +2)^2 + 4a\omega m - 4a^2 \omega^2} \\ \nn
                & \times \lrsbrk{\lambda^2+36a\omega m -36 a^2 \omega^2} \\ \nn
                & + (2\lambda +3)(96a^2\omega^2 - 48a\omega m) + 144\omega^2(1-a^2)\,,
 \end{align} 
 with 
 \begin{align}
 \text{Im} \, C &= 12\omega\,, \\ 
 \text{Re} \, C &= +\sqrt{\vert C\vert^2-(\text{Im} \, C)^2} \,. 
 \end{align}
Here we have defined
\begin{equation}
\label{eq:epsilon-def}
\varepsilon  =   \frac{\sqrt{1-a^2}}{4r_H}\,,  
\end{equation}
and $\lambda \equiv {}_{-2}\lambda_{\lmo}$ is the eigenvalue of the $s=-2$ spin-weighted spheroidal harmonic.  See Appendix.~\ref{app:conservation-erg}  for more discussions on the Teukolsky-Starobinsky identity.

Combining Eq.~\eqref{generalbc} with Eq.~\eqref{eq:Y-Z-relation}, we finally arrive at the relation between $Z^{\rm refl}$ and $ Z^{\rm in}$, which is given by
\begin{equation}
\label{eq:Zrefl-to-Zhole}
Z^{\rm refl}_{\lmo} = \sum_{\ell^\prime} \mathcal{G}_{\ell \ell^\prime m \omega }  Z^{\rm hole^*}_{\lpmmmos}\,,
\end{equation}
where 
\begin{equation}
\mathcal{G}_{\ell \ell^\prime m \omega } = (-1)^{m+1}\frac{1}{4} e^{-2ikb_*} \mathcal{M}_{\ell \ell^\prime m \omega } \frac{D_{\lpmo}}{C_{\lpmo}} \,.
\end{equation}
We have used the relations $D_{\lmo} = D^*_{\lmmmos}$ and $C_{\lmo} = C^*_{\lmmmos}$ in order to obtain the above equation.   

If we restrict ourselves to the simple case where $\ell$- and $\ell^\prime$- modes do not mix up, we may simply write Eq.~\eqref{eq:Zrefl-to-Zhole} as
\begin{align}
\label{eq:reflect-out-in}
Z^{\rm refl}_{\lmo} & = \hat{\mathcal{G}}_{\lmo}   \, Z^{\rm hole^*}_{\lmmmo} \,, 
\end{align}
with 
\begin{align}
\label{eq:K-expression}
\hat{\mathcal{G}}_{\lmo} &\equiv  \frac{D_{\lmo}}{4 C_{\lmo}} \, \mathcal{R}^*_{\text{-}\omega+ m \Omega_H}\, e^{i \varphi^{\rm refl}_{\lmo}}\,,
\end{align}
and 
\begin{equation}
\label{eq:ref-phase}
\varphi^{\rm refl}_{\lmo} = (m+1)\pi-2 k b_* \,.
\end{equation}
Eq.~\eqref{eq:reflect-out-in} says that, \textit{the $(\ell,\,m, \,\omega)$-mode of gravitational-wave echoes are not induced by the reflection of the incoming $(\ell,\,m, \,\omega)$-mode but the $(\ell,\,-m, \,-\omega^*)$-mode instead}.  The mixing of these two types of modes essentially indicates the breaking of isospectrality as pointed out by Ref.~\cite{Maggio:2020jml}.  We will get back to this point later.
The other new result is the extra phase term $\varphi^{\rm refl}$ for the reflected waves, which may be important for observations.

\subsection{Reflectivity for Sasaki-Nakamura Mode Amplitudes}

Since most previous literatures on gravitational wave echoes base their models on the reflection of  Sasaki-Nakamura (SN) functions, one may ask how the tidal reflectivity can be related to the SN reflectivity. (See Appendix~\ref{app:Teukolsky-eq} for a brief review of the SN formalism.)  In the vicinity of the horizon, the $s=-2$ SN function, i.e., the one {\it associated with} $\psi_4$, can be written as 
\begin{align}
\label{eq:R-under-CSN}
X^{\rm ECO}_{\lmo} & =  \xi ^{\rm hole}_{\lmo} \, e^{-ikr_*} +  {\xi^{\rm refl}_{\lmo}}\, e^{ikr_*} \,. & r_* \rightarrow b_* \,.
\end{align}
Under the Chandrasekhar-Sasaki-Nakamura transformation, we have
\begin{equation}
\label{CSN}
 \xi ^{\rm hole}_{\lmo} =Z^{\rm hole}_{\lmo} d_{\lmo}\,,\quad
  {\xi^{\rm refl}_{\lmo}} = \frac{Z^{\rm refl}_{\lmo}}{f_{\lmo}} 
  \end{equation}
with
\begin{align}
d_{\lmo} = \sqrt{2 r_H} [ &(8-24i\omega -16\omega^2) r^2_H \nn \\
 & +  (12iam-16+16am\omega+24i\omega)r_H  \nn \\
& - 4a^2m^2-12iam+8]\,, 
\end{align}
and 
\begin{align} 
f_{\lmo} = - \frac{4k\sqrt{2r_H} \lrsbrk{2kr_H + i(r_H-1)}}{\eta(r_H)} \,.
\end{align}
Inserting Eqs.~\eqref{CSN} into  Eq.~\eqref{eq:Zrefl-to-Zhole}, we obtain boundary condition for the $\lmo$ components of the SN functions: 
\begin{equation}
\xi_{\lmo}^{\rm refl} =  \frac{(-1)^{m+1} e^{-2ikb_*}}{4 f_{\lmo}}\sum_{\ell^\prime} \mathcal{M}_{\ell \ell^\prime m \omega } \frac{D_{\lpmo} }{C_{\lpmo} d_{\lpmo}} \xi^{\rm hole^*}_{\lpmmmos}\,,
\end{equation}
Here we have used the identity that
\begin{equation}
d_{\lmo} = d_{\lmmmos}^*\,.
\end{equation}
As we will see later, the fact that reflection at the ECO surface turns the in-going $(\ell,-m,-\omega)$ SN components into out-going $(l,m,\omega)$ SN components leads to the breaking of isospectrality, which has also been pointed out by Maggio {\it et al.}~\cite{Maggio:2020jml}; here we take the further step of relating these coefficients to the tidal response of the ECO.

\begin{figure*}[th]
\centering
\includegraphics[width=0.48\textwidth]{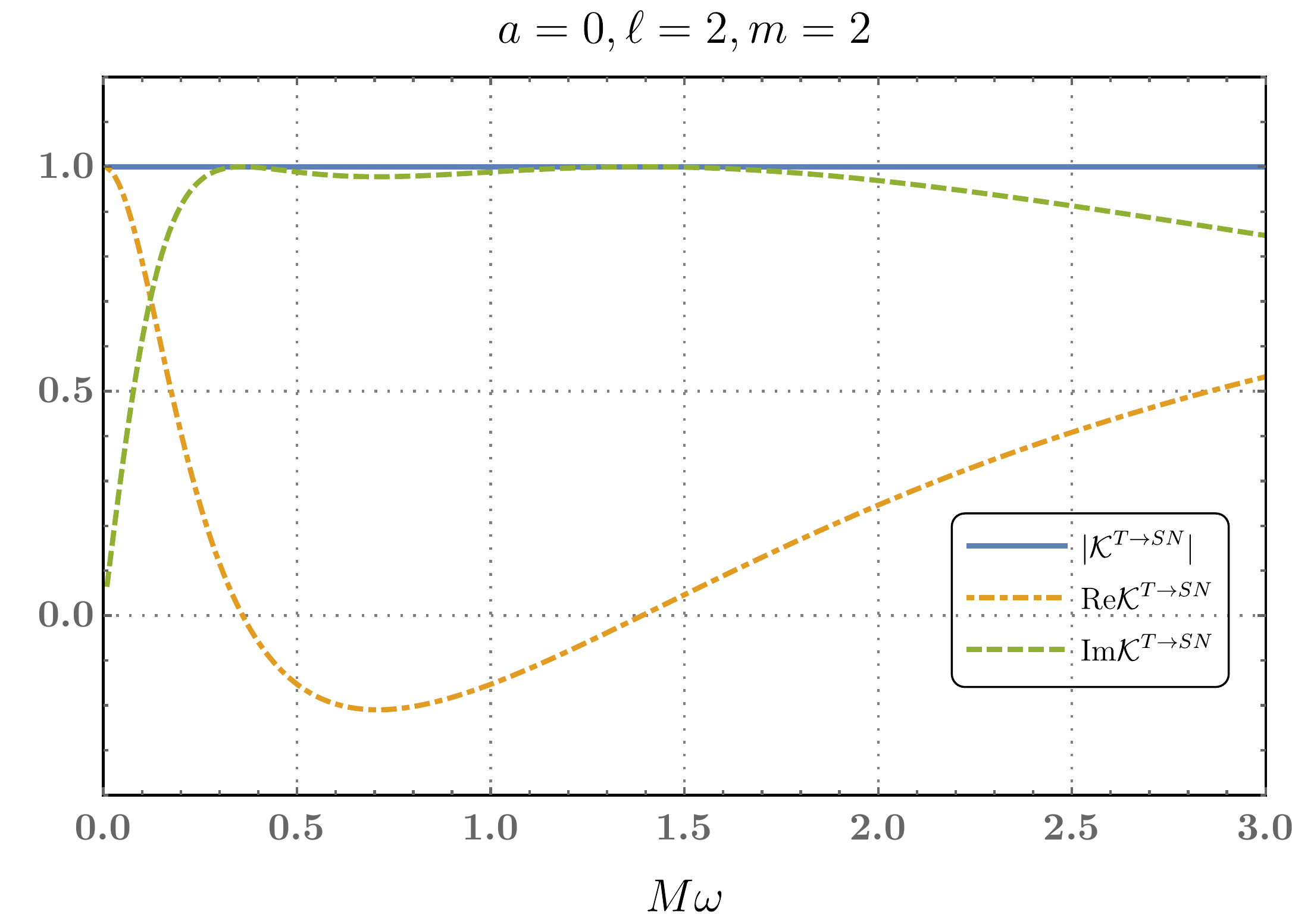}
\includegraphics[width=0.48\textwidth]{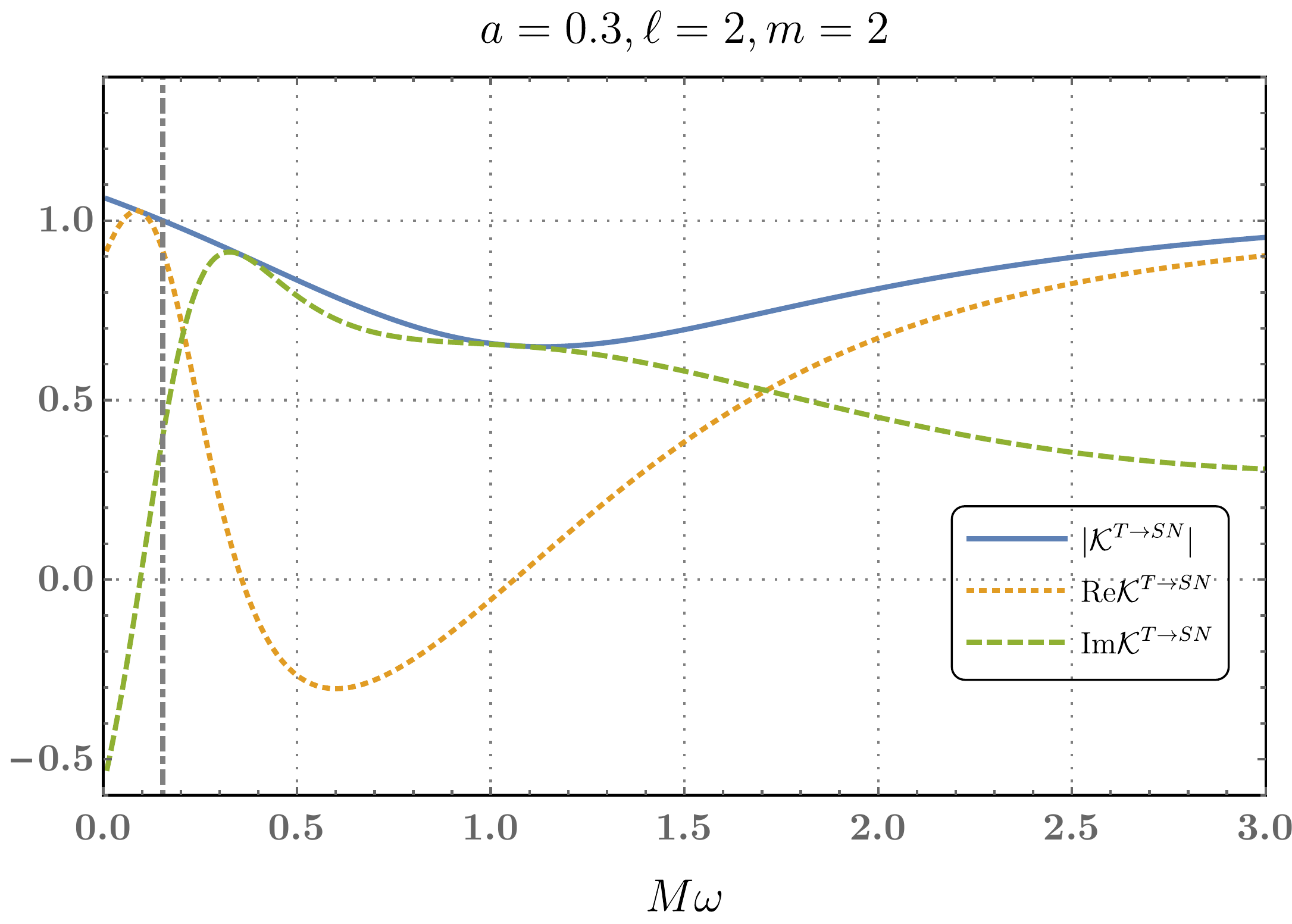}
\includegraphics[width=0.48\textwidth]{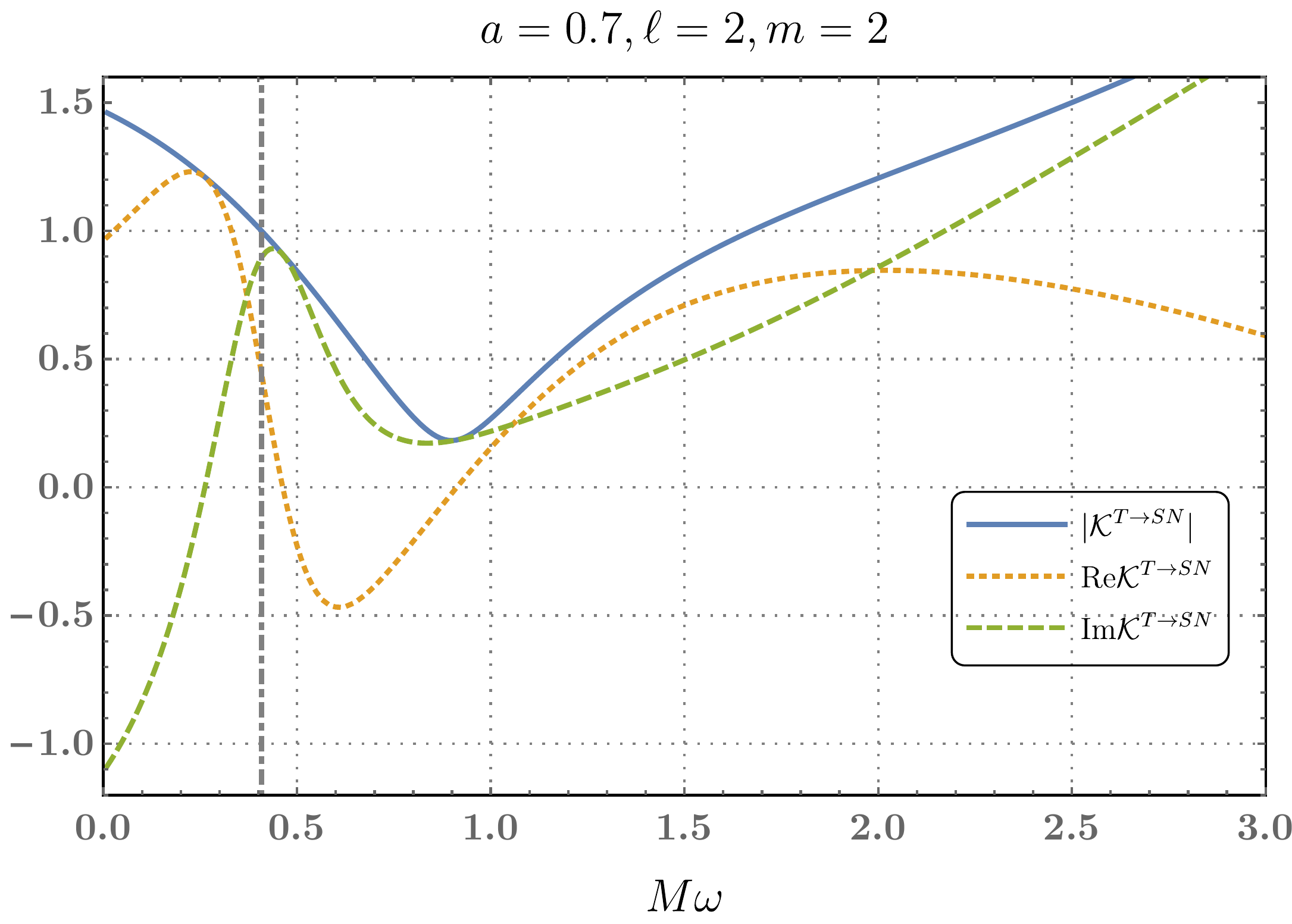}
\includegraphics[width=0.48\textwidth]{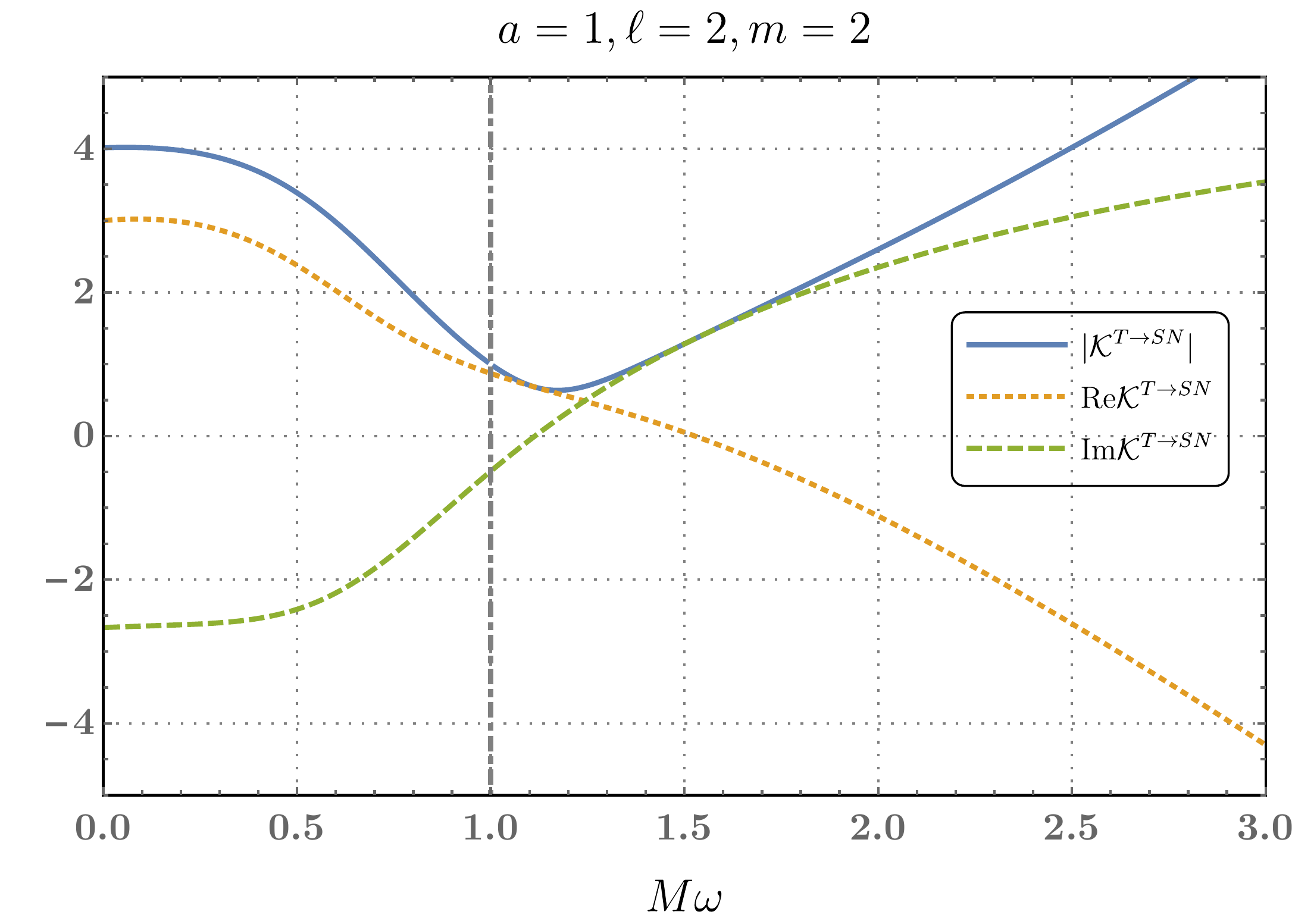}
\caption{Conversion factor $ \mathcal{K}_{\lmo}^{\rm T\rightarrow SN}$ from the Teukolsky $\mathcal{R}$ to the Sasaki-Nakamura $\mathcal{R}_{\mathrm{SN}}$. Here we have ignored the $\ell$-$\ell'$ mode mixing.  We plot  real and imaginary parts, as well as the modulus, of $\mathcal{K}_{\lmo}^{\rm T\rightarrow SN}$ for the $(2, 2)$-mode with $a=0, 0.3, 0.7, 1$ respectively.  The gray dot-dashed line marks the horizon frequency $m\Omega_H$. 
In the Schwarzschild case, the two reflection coefficients only differ by a phase.  For Kerr spacetimes we have $ |\mathcal{K}_{\lmo}^{\rm T\rightarrow SN}| >1$  for both low and high frequencies, but $| \mathcal{K}_{\lmo}^{\rm T\rightarrow SN}|$ dips below 1 for some frequencies.  Also note that $\vert \mathcal{K}_{\lmo}^{\rm T\rightarrow SN}\vert=1$ when $\omega$ equals to the horizon frequency.}
\label{fig:R-to-R-factor}
\end{figure*}

 For the most simplified scenario where $Z^{\rm hole^*}_{\lmmmos}=Z^{\rm hole}_{\lmo}$ and different $\ell^\prime$-modes do not mix, we may simply write 
 \begin{equation}
 \xi_{\lmo}^{\rm refl} = \mathcal{R}_{lmo}^{\rm SN}\xi^{\rm hole}_{\lmo}\,,
\end{equation}
where
\begin{equation}
\label{eq:RSN-R-relation}
\mathcal{R}^{\rm SN}_{\lmo} = \mathcal{K}_{\lmo}^{\rm T\rightarrow SN} \mathcal{R}^*_{\text{-}\omega+m\Omega_H} \,,
\end{equation}
with 
\begin{equation}
 \mathcal{K}_{\lmo}^{\rm T\rightarrow SN}= \frac{(-1)^{m+1}D_{\lmo}}{4C_{\lmo}f_{\lmo}d_{\lmo}}\,.
\end{equation}
This is a simple linear factor that converts $\mathcal{R}$ into the $\mathcal{R}^{\rm SN}$ that are used in  SN calculations.  In the Schwarzschild limit, we have 
\begin{equation}
\mathcal{K}_{\lmo}^{\rm T\rightarrow SN} =\frac{(-1)^m (4 \omega -i) \left[12 \omega + i \lambda( \lambda +2)  \right]}{(4 \omega +i) \left[12 \omega -i \lambda  (\lambda +2)\right]}\,,\quad a=0\,,
 \end{equation}
 where $\lambda = (\ell-1)(\ell+2)$.  One immediately notes that $\vert \mathcal{K}_{\lmo}^{\rm T\rightarrow SN} \vert=1$ in the Schwarzschild limit.  For spinning ECOs,
we numerically investigate $ \mathcal{K}_{\lmo}^{\rm T\rightarrow SN}$ for the $(2,2)$ mode for different spins in Fig.~\ref{fig:R-to-R-factor}.



%

\subsection{Energy Contents of Incoming and Reflected Waves}

The reflection coefficient we defined in last subsection is indeed the (square root of) power reflectivity of the gravitational waves on the ECO boundary.  To see this, consider a solution to $s=-2$ Teukolsky equation near the ECO surface.  The energy flux down to the surface is given by~\cite{Teukolsky:1974yv}
\begin{equation}
\frac{d E_{\rm hole}}{d\omega}  = \sum_{\lm} \frac{\omega}{64\pi k(k^2+4\epsilon^2) (2r_H)^3} \vert Y^{\rm hole}_{\lmo} \vert^2 \,,
\end{equation} 
while the energy propagating outward from the surface is given by
\begin{equation}
\frac{d E^{\rm refl}}{d \omega} = \sum_{\lm} \frac{\omega}{4\pi k (k^2+4\epsilon^2)(2r_H)^3} \vert Z^{\rm refl}_{\lmo} \vert^2\,.
\end{equation}
Here $\omega$ are all taken as real numbers.  See Appendix~\ref{app:conservation-erg} for detailed discussions on the energy flux and the energy conservation law.  In the simple case of neglecting $\ell$-$\ell'$ mixing, incoming energy from the $(\lmmmo)$-mode will return from the $(\lmo)$-mode, with 
\begin{equation}
\left(\frac{dE_{\rm refl}}{d\omega}\right)_{\lmo} =\vert \mathcal{R}_{\text{-}\omega+m\Omega_H} \vert^2
\left(\frac{dE_{\rm hole}}{d\omega}\right)_{\lmmmo} 
\end{equation}
This means our reflectivity $\mathcal{R}$ indeed acts as an energy reflectivity.

\comments{
For each mode, the power reflection rate on the ECO boundary is then given by the ratio of outgoing energy to ingoing energy, which yields
\begin{align}
\lrbrk{\frac{E^{\rm refl}_{\lmo}}{E^{\rm hole}_{\lmo}}}_{r\rightarrow b} = \frac{16 \,\vert C_{\lmo}\vert^2} {\vert D_{\lmo} \vert^2}\frac{\vert Z^{\rm refl}_{\lmo} \vert^2}{\vert Z^{\rm hole}_{\lmo}\vert^2} \,.
\end{align}
Plugging in Eq.~\eqref{eq:reflect-out-in},~\eqref{eq:K-expression} and~\eqref{eq:ref-phase}, one immediately finds that 
\begin{equation}
\lrbrk{\frac{E^{\rm refl}_{\lmo}}{E^{\rm hole}_{\lmo}}}_{r \rightarrow b}= \vert \mathcal{R}_{\text{-}\omega+m\Omega_H} \vert^2  \frac{\left\vert Z^{\rm hole}_{\lmmmo} \right\vert^2}{\left\vert Z^{\rm hole}_{\lmo}\right\vert^2}\,.
\end{equation}
This means our reflectivity $\mathcal{R}$ is indeed the energy reflectivity which turns an incoming $(\ell, m, \omega)$-mode into an outgoing $(\ell, -m, -\omega)$-mode.
 }

\section{Waveforms and Quasi-Normal Modes of the ECO}
\label{sec:waveform-qnm}

In this section, we show how our ECO boundary conditions can be applied to echo computations and resonant conditions for quasi-normal modes.  We shall also restrict ourselves to the case of 
\begin{equation}
\hat{\mathcal{G}}_{\lmo} =\hat{\mathcal{G}}^*_{\lmmmos}\,.
\end{equation}
This is satisfied by all the reflectivity models discussed in this paper, since in these cases the tidal response in the time-domain, $\mathcal{R}(b,\theta;t)$ [Cf.~\eqref{eq:ref-model}]  is real-valued.

\subsection{Even and Odd-Parity Echoes}

In this subsection, we derive the gravitational-wave echo waveform based on our reflection model.  Note that this echo can be the additional wave due to the reflection at the ECO surface during the inspiral phase --- it does not necessarily has to be the echo that follows the ringdown phase of the coalescence wave. 

Suppose now, we have some small perturbations towards the ECO spacetime.  We assume that the source in the Teukolsky equation drives a ${}_{-2}\Upsilon^{(0)}$, which has the following form at $r_*\rightarrow -\infty$: 
\begin{align}
{}_{-2}\Upsilon^{\rm (0)} 
=  \sum_{\lm} \int\frac{d\omega}{2\pi} 
Z_{\lmo}^{\rm hole\,(0)} \Delta^2 e^{-ikr_*}
  {}_{-2}S_{\lmo} (\theta,\phi)e^{-i\omega t} \,.
\end{align}
This satisfies the Teukolsky equation with the appropriate source term away from the horizon, the out-going condition at infinity, but not the ECO boundary condition near the horizon.  We will need to add an additional homogeneous solution, which satisfies the out-going boundary condition at infinity.  Recall that for the radial part, we have
\begin{align}
{R}^{+\infty}_{\lmo} = \left\{ 
\begin{array}{ll}
 D^{\rm in}_{\lmo}\Delta^{2} e^{-i k r_*} +  D^{\rm out}_{\lmo} e^{i k r_*} \,,  & r \rightarrow b \,, \\ \nn
 \\ \nn
  r^3 e^{i \omega r_*} \,,  & r \rightarrow +\infty \,.
 \end{array}\right.
\end{align}   
Thus we add the following homogeneous solution to $\Upsilon^{\rm (0)} $: 
\begin{equation}
{}_{-2}\Upsilon^{\rm echo}  = \sum_{\lm}\int\frac{d\omega}{2\pi} c_{\lmo} R_{\lmo}^{+\infty} {}_{-2}S_{\lmo} (\theta,\phi)e^{-i\omega t} \,,
\end{equation} 
so that ${}_{-2}\Upsilon^{(0)}+{}_{-2}\Upsilon^{\rm echo}$ is of the form \eqref{eq:Teukolsky-H-decompose}, also satisfying \eqref{eq:reflect-out-in}.   The asymptotic behavior of ${}_{-2}\Upsilon^{\rm echo}$ is given by 
\begin{widetext}
\begin{equation}
{}_{-2}\Upsilon^{\rm echo}  = 
\left\{
\begin{array}{ll}
\displaystyle \sum_{\lm}\int \frac{d\omega}{2\pi} c_{\lmo}  r^3 e^{+i\omega r_*} e^{-i\omega t}  {}_{-2}S_{\lmo} (\theta,\phi) \,, \quad  & r_*\rightarrow +\infty\,, \\
\\
\displaystyle  \sum_{\lm}\int\frac{d\omega}{2\pi} c_{\lmo} \left[ D^{\rm in}_{\lmo}\Delta^{2} e^{-i k r_*} +  D^{\rm out}_{\lmo} e^{i k r_*}\right]  e^{-i\omega t}  {}_{-2}S_{\lmo} (\theta,\phi) \,, \quad  & r_*\rightarrow b_* \,.\\
\end{array}\right.
\end{equation} 
Here we already see that the amplitudes $c_{lm\omega}$  directly give us the {\it additional} gravitational waves due to the reflecting surface.  Identifying term by term between ${}_{-2}\Upsilon^{(0)}+{}_{-2}\Upsilon^{\rm echo}$  and Eq.~\eqref{eq:Teukolsky-H-decompose}, we find
\begin{equation}
Z^{\rm hole}_{\lmo} = Z^{\rm hole\,(0)} _{\lmo}+ c_{\lmo} D^{\rm in}_{\lmo} \,, \quad
Z^{\rm refl}_{\lmo} = c_{\lmo} D^{\rm out}_{\lmo} \,.
\end{equation}
Applying Eq.~\eqref{eq:Zrefl-to-Zhole}, we obtain
\begin{align}
c_{\lmo}D^{\rm out}_{\lmo}  &= \sum_{\ell^\prime} \mathcal{G}_{\ell \ell^\prime m \omega} 
\left[ Z^{\rm hole\,(0)*} _{\lpmmmo}+ c^*_{\lpmmmo} D^{\rm in\,*}_{\lpmmmo}\right] \,, \nonumber\\
c_{\lmmmo}^* D^{\rm out\,*}_{\lmmmo}  &= \sum_{\ell^\prime} \mathcal{G}^*_{\ell \ell^\prime \text{-}m \text{-}\omega} 
\left[ Z^{\rm hole\,(0)} _{\lpmo}+ c_{\lpmo} D^{\rm in}_{\lpmo}\right]\,.
\end{align}
Here we restrict ourselves to real-valued $\omega$ only.   Using the symmetry of the Teukolsky equation, for real-valued $\omega$, it is straightforward to show that the homogeneous solutions have the symmetry that  
\begin{equation}
D^{\rm in}_{\lmo} =D^{\rm in\,*}_{\lmmmo} \,,\quad 
D^{\rm out}_{\lmo} =D^{\rm out\,*}_{\lmmmo}\,.
\end{equation}
We can then write
\begin{align}
\label{inoutmatrix}
\left(
\begin{array}{cc}
\delta_{\ell\ell^\prime}D^{\rm out}_{\lmo}  & -\mathcal{G}_{\ell\lpmo}D^{\rm in}_{\lmo} \\
-  \overline{\mathcal{G}}_{\ell\lpmo} D^{\rm in}_{\lmo} & \delta_{\ell\ell^\prime} D^{\rm out}_{\lmo} 
\end{array}
\right) &
\left(
\begin{array}{c}
c_{\lpmo} \\
c_{\lpmmmo}^* 
\end{array}\right)
= 
\left(
\begin{array}{cc}
\mathcal{G}_{\ell\lpmo} & 0 \\
0 & \overline{\mathcal{G}}_{\ell\lpmo}
\end{array}
\right)
\left(
\begin{array}{c}
Z^{\rm in\,(0)*} _{\lpmmmo} \\
Z^{\rm in\,(0)} _{\lpmo}
\end{array}
\right)\,, \\ 
&\overline{\mathcal{G}}_{\ell\lpmo} \equiv \mathcal{G}_{\ell\lpmmmo}^*\,, 
\end{align}
\end{widetext}
where the components in all matrices are also block matrices with $\ell$ and $\ell^\prime$ representing sections of rows and columns.
This will allow us to solve for $c_{\lmo}$, therefore leading to the additional out-going waves at infinity, i.e.  the gravitational-wave echoes. 

In the simple case where there is no  $\ell$-$\ell'$ mixing for reflected waves (so that the relation between reflected waves and incoming waves is simply given by Eq.~\eqref{eq:reflect-out-in}), and that 
\begin{equation}
\hat{\mathcal{G}^*}_{\lmmmo} \equiv \hat{\mathcal{G}}_{\lmo}\,,
\end{equation}
we can have simpler results.  
For each harmonic for the $Z$ components (similar for the $c$ components), we can define symmetric and anti-symmetric quadrature amplitudes
\begin{align}
Z^{\rm hole\,(0),S}_{\lmo} &\equiv 
\frac{Z^{\rm hole\,(0)}_{\lmo}+Z^{\rm hole\,(0)\,*}_{\lmmmo}}{\sqrt{2}}
\,,\\ 
Z^{\rm hole\,(0),A}_{\lmo} &\equiv 
\frac{Z^{\rm hole\,(0)}_{\lmo}-Z^{\rm hole\,(0)\,*}_{\lmmmo}}{\sqrt{2}i}\,.
\end{align}
We then have
\begin{align}
c_{\lmo}^{\rm S}  & =\frac{\hat{\mathcal{G}}_{\lmo}}{D_{\lmo}^{\rm out} - \hat{\mathcal{G}}_{\lmo} D_{\lmo}^{\rm in}} Z_{\lmo}^{\rm hole\,(0),S} \,,\\
c_{\lmo}^{\rm A}  & =-\frac{\hat{\mathcal{G}}_{\lmo}}{D_{\lmo}^{\rm out} + \hat{\mathcal{G}}_{\lmo} D_{\lmo}^{\rm in}} Z_{\lmo}^{\rm hole\,(0),A}\,.
\end{align}
Here we see that the $A$ quadrature has a reflectivity of $-\hat{\mathcal{G}}_{\lmo}$, compared with $\hat{\mathcal{G}}_{\lmo}$ for the $S$ quadrature.  These quadratures correspond to electric- and magnetic-type perturbations. 

As it turns out, non-spinning binaries, or those with spins aligned with the orbital angular momentum, only excite the $S$ quadrature --- although generically both quadratures are excited --- they will have different ech{}oes.  In the case when echoes are well-separated in the time domain, the first, third, and other odd echoes, the $A$ and $S$ will have transfer functions negative to each other, while for even echoes, they will have the same transfer function. 

If we further simplify the problem by demanding $c_{\lmo} = c^*_{\lmmmo}$, Eq.~\eqref{inoutmatrix} gives that
\begin{equation}
c_{\lmo} = \frac{ \hat{\mathcal{G}}_{\lmo}  }{D^{\rm out}_{\lmo} - \hat{\mathcal{G}}_{\lmo}  \,  D^{\rm in}_{\lmo} } Z^{\rm hole \, (0)}_{\lmo} \,.
\end{equation}
This expression coincides, for instance, with the one obtained in~\cite{Mark:2017dnq} for the spherically-symmetric spacetime with a reflecting surface. Note that the phase factor $e^{-2ik b_*}$ has been absorbed into our definition of $\hat{\mathcal{G}}$.

\subsection{Echoes driven by symmetric source terms}

In our reflection model \eqref{eq:reflect-out-in},
as discussed in Ref.~\cite{Hughes:1999bq}, the coefficients $Z^{\rm hole^*}_{\lmmmos}$ and $Z^{\rm hole}_{\lmo}$ are related for quasi-circular orbits.  For such orbits, one  can define a series of frequencies as
\begin{equation}
\omega_{mk} = m \Omega_\phi + k \Omega_\theta\,,
\end{equation}
where $\Omega_\phi$ and $\Omega_\theta$ are two fundamental frequencies defined for periodic motions in $\phi$ and $\theta$.  Then, for real frequencies, we can decompose the amplitude $Z^{\rm in}_{\lmo}$ according to
\begin{equation}
Z^{\rm hole}_{\lmo} = \sum_k Z^{\rm hole}_{\lm k} \delta (\omega -\omega_{mk})\,.
\end{equation}
It is easy to check that for Kerr black holes,
\begin{equation}
Z^{\rm hole^*}_{\ell \text{-}m \text{-}k} = (-1)^{\ell+k} Z^{\rm hole}_{\lm k} \,.
\end{equation}
That is, if we consider a specific circular orbit, we have the symmetry that $Z^{\rm hole^*}_{\lmmmos}$ is either equal to $Z^{\rm hole}_{\lmo}$, or they differ by a minus sign.  In this simple case, our reflection model~\eqref{eq:reflect-out-in} does not involve different modes, and the model becomes similar to those reflection models based on Sasaki-Nakamura functions like in Ref.~\cite{Wang:2019rcf}.  However, if we consider the full quasi-circular motions, i.e. adding up all orbits, this symmetry no longer exists, and one has to consider the mixing of modes when dealing with the reflecting boundary.  For general orbits that are not quasi-circular, the symmetry between $Z^{\rm hole^*}_{\lmmmos}$ and $Z^{\rm hole}_{\lmo}$ may not exist.

Now for the symmetric source, where there is no mode mixing, let us consider a solution ${}_{-2}\Upsilon^{(0)}$ to the Teukolsky equation, which has the following form at $r_* \rightarrow -\infty$: 
\begin{align}
{}_{-2}\Upsilon^{\rm (0)} 
=  \sum_{\lm} \int\frac{d\omega}{2\pi} 
Z_{\lmo}^{\rm hole\,(0)} \Delta^2 e^{-ikr_*}
  {}_{-2}S_{\lmo} (\theta,\phi)e^{-i\omega t} \,.
\end{align}
Following the same steps as in the last subsection, it is straightforward to show that the echo solution to the Teukolsky equation at infinity is given by
\begin{equation}
{}_{-2}\Upsilon^{\rm echo}  = \sum_{\lm}\int\frac{d\omega}{2\pi} Z^{\rm echo}_{\lmo} {}_{-2}S_{\lmo} (\theta,\phi)e^{-i\omega t} \,,
\end{equation}  
with
\begin{equation}
\label{eq:echo-tidal-simple}
Z^{\rm echo}_{\lmo} = \frac{ \hat{\mathcal{G}}_{\lmo}  }{D^{\rm out}_{\lmo} - \hat{\mathcal{G}}_{\lmo}  \,  D^{\rm in}_{\lmo} } Z^{\rm hole \, (0)}_{\lmo} \,,
\end{equation}
where we have chosen the normalization $D^{\infty}_{\lmo} =1$.  The tidal reflectivity can also be directly related to the SN reflectivity as 
\begin{equation}
\label{eq:RSN-R-relation}
\mathcal{R}^{\rm SN}_{\lmo} = (-1)^{m+1}\frac{D_{\lmo}}{4C_{\lmo}f_{\lmo}d_{\lmo}} \mathcal{R}^*_{\text{-}\omega+m\Omega_H}\,.
\end{equation}
In this simple scenario, the tidal reflectivity is exactly the energy reflectivity for each mode.

\subsection{Quasi-Normal Modes and Breakdown of Isospectrality}

For Quasi-Normal Modes, 
we set $Z$ to zero, and analytically continue Eq.~\eqref{inoutmatrix} to complex $\omega$. The QNM frequencies can be directly solved by setting the determinant of the lhs matrix of Eq.~\eqref{inoutmatrix} to zero, i.e.
\begin{equation}
\text{det}\left(
\begin{array}{cc}
\delta_{\ell\ell^\prime}D^{\rm out}_{\lmo}  & -\mathcal{G}_{\ell\lpmo}D^{\rm in}_{\lmo} \\
-  \mathcal{G}_{\ell\lpmmmos}^* D^{\rm in}_{\lmo} & \delta_{\ell\ell^\prime} D^{\rm out}_{\lmo} 
\end{array}
\right) = 0\,.
\end{equation}
This will in general cause a mixing between QNMs with different $\ell$, and break the {\it isospectrality property} of the Kerr spacetime and lead to {\it two distinct QNMs} for each $(\ell,m)$.

Neglecting the $\ell$-$\ell^\prime$ mixing, we can simply write
\begin{align}
\left[D^{\rm out}_{\lmo}\right]^2 & = \hat{\mathcal{G}}_{\lmo} \bar{\hat{\mathcal{G}}}_{\lmo}
\left[D^{\rm in}_{ \lmo}\right]^2 \,, & \bar{\hat{\mathcal{G}}}_{\lmo} \equiv \hat{\mathcal{G}^*}_{\lmmmos}\,. 
\end{align}
In the special case of $\hat{\mathcal{G}}_{\lmo}= \bar{\hat{\mathcal{G}}}_{\lmo}$ (which is satisfied by all the reflectivity models discussed in this paper), we note that the ECO's QNMs split into $S$ and $A$ modes, with $\omega_{n\lm}^S$ and $\omega_{n\lm}^A$ satisfying different equations:  
\begin{align}
D_{\lmo_S}^{\rm out} - \hat{\mathcal{G}}_{\lmo_S} D_{\lmo_S}^{\rm in} &=0\,, \\
D_{\lmo_A}^{\rm out} + \hat{\mathcal{G}}_{\lmo_A} D_{\lmo_A}^{\rm in} &=0\,.
\end{align}
This still breaks the  { isospectrality} properties of Kerr spacetime. Note that this property has also been found and studied in Ref.~\cite{Maggio:2020jml} with their echo model which describes the ECO as a dissipative fluid.   Since modes of the ECO are usually excited collectively, the main signature of the breakdown of isospectrality is still the fact that $S$ and $A$ echoes have alternating sign differences in even and odd echoes.

\section{Conclusions}
\label{sec:summary}

In this paper, we developed a more physical way to impose boundary conditions for Teukolsky functions near the surface of extremely compact objects.   We adopted the Membrane Paradigm, and assumed that the ECO structure is well adapted to the coordinate system of the Fiducial Observers, which is an approximate Rindler coordinate system near the horizon.  More specifically, assuming that the additional physics near an ECO can be viewed as modified propagation laws of gravitational waves in the Rindler coordinate system, we were able to obtain reflectivity models for spinning ECOs that are similar to those proposed by previous literature, when taking the Schwarzschild limit.  In particular, the Boltzmann reflectivity of Oshita {\it et al.} was obtainable from a position-dependent damping of gravitational waves in the Rindler coordinate system, which might be thought of  as due to the emergent nature of gravity. 

As it has turned out, the most directly physical condition is between in-going components of $\psi_0$ and out-going components of $\psi_4$, although relations between in-going and out-going components of $\psi_4$, as well as those of the Sasaki-Nakamura functions, can be obtained by using the Starobinsky-Teukolsky transformation, as well as the Chandrasekhar-Sasaki-Nakamura relations.  

The deformation of space-time geometry due to the spin of the ECO causes a mixing between different $\ell$ modes during reflection at the ECO surface; reflection at the ECO also takes $(m,\omega) \rightarrow(-m,-\omega^*)$.  This means an incoming $(\ell, m,\omega)$ mode is reflected into $(\ell',-m,-\omega^*)$ modes.  For moderately rapidly spinning holes, such $\ell$-$\ell'$ mixing is moderate, but non-negligible, which means accurately modeling echoes will indeed have to take such mixing into account.  For incoming waves toward the ECO caused by a quasi-circular inspiral of a non-spinnining particle, the waveform has a {\it definite partiy}, and is invariant under the $(m,\omega) \rightarrow(-m,-\omega^*)$ transformation.  For more general waves, the  $(m,\omega) \rightarrow(-m,-\omega^*)$ map causes echoes from even- and odd-parity waves to differ from each other; it also causes the breakdown of quasi-normal mode isospectrality, as has been pointed out by Maggio {\it et al.} in the Schwarzschild case.

\acknowledgments
The authors would like to thank
Shuo Xin,
Wenbiao Han,
Ka-Lok R. Lo,
Ling Sun
and 
Niayesh Afshordi
for useful conversations.  BC and YC acknowledge the support from the Brinson Foundation, the Simons Foundation (Award Number 568762), and the National Science Foundation, Grants PHY-2011961, PHY-2011968. 
QW acknowledges the support from the University of Waterloo, Natural Sciences and Engineering Research Council of Canada (NSERC), and the Perimeter Institute for Theoretical Physics. \\
\appendix

\section{The homogeneous Teukolsky and Sasaki-Nakamura equations}
\label{app:Teukolsky-eq}
Perturbations of Kerr spacetime can be described by the Teukolsky equations~\cite{Teukolsky:1973ha}.   In the vacuum case, one can decompose solutions to the homogeneous Teukolsky equation as 
\begin{align}
{}_{s}\Upsilon = \sum_{\lm}\int \frac{ d \omega}{2\pi} \, e^{-i\omega t+ i m\phi}\,{}_{s}R_{\lmo}(r) {}_{s}S_{\lmo}(\theta) \,,
\end{align}  
where ${}_{s}S_{\lmo}(\theta)$ is the spin-weighted spheroidal harmonic function, and $s$ is the spin weight.
The Teukolsky equations are then separable, and the equations for $R$ and $S$ are respectively
\begin{widetext}
\begin{align}
\label{eq:Teukolsky-radial}
&\lrsbrk{\Delta^{-s} \frac{d}{d r} \lrbrk{\Delta^{s+1} \frac{d }{d r} } + \frac{K^2-2is\lrbrk{r-1}K}{\Delta} + 4is\omega r -{}_s\lambda_{\lmo} }{}_{s}R_{\lmo} = 0 \,, \\
&\lrsbrk{\frac{1}{\sin\theta} \frac{d }{d\theta} \lrbrk{\sin\theta \frac{d }{d\theta}} - a^2\omega^2 \sin^2\theta - \frac{(m+s\cos\theta)^2}{\sin^2\theta} -2a\omega s\cos\theta + s + 2 m a\omega + {}_s\lambda_{\lmo} }{}_{s}S_{\lmo} = 0 \,,
\end{align}
\end{widetext}
where $K = (r^2 + a^2)\omega -m a$, and ${}_s\lambda_{\lmo}$ is the eigenvalue of the spin-weighted spheroidal harmonic.

For $s=-2$, the radial equation~\eqref{eq:Teukolsky-radial} admits two independent solutions, ${}_{-2}R^{\rm H}_{\lmo}$ and ${}_{-2}R^{\infty}_{\lmo}$, which have the following asymptotic forms:
\begin{align}
{}_{-2}R^{\rm H}_{\lmo} &= \left\{
\begin{array}{lr}
B^{\rm out}_{\lmo} r^3 e^{i\omega r_*} + B^{\rm in}_{\lmo}r^{-1} e^{-i\omega r_*} \,, & r\rightarrow \infty\,, \\
\\
B^{\rm hole}_{\lmo} \Delta^2 e^{-ik r_*} \,, & r\rightarrow r_H \,; 
\end{array} \right.\\ 
{}_{-2}R^{\rm \infty}_{\lmo} &= \left\{
\begin{array}{lr}
D^{\infty}_{\lmo} r^3 e^{i\omega r_*} \,, & r\rightarrow \infty\,, \\
\\
D^{\rm out}_{\lmo} e^{ik r_*} + D^{\rm in}_{\lmo} \Delta^2 e^{ -ik r_*} \,, & r\rightarrow r_H \,.
\end{array} \right.
\end{align}

The Sasaki-Nakamura-Chandrashekar transformation~\cite{Sasaki:1981sx} takes the Teukolsky radial function ${}_{-2}R(r)$ to the Sasaki-Nakamura function $X(r)$, and the Teukolsky equation becomes the Sasaki-Nakamura equation.  The homogeneous SN equation is given by
\begin{equation}
\frac{d^2 X_{\lmo}}{d r^2_*} - F(r) \frac{d X_{\lmo}}{d r_*} - U(r) X_{\lmo} = 0 \,.
\end{equation}
The explicit expressions for $F(r)$ and $U(r)$ are given in Ref.~\cite{Sasaki:2003xr}'s  Eqs.~(51-58).  
The SN equation also admits two independent solutions, $X^{\rm H}_{\lmo}$ and $X^{\infty}_{\lmo}$, which have the asymptotic values:
\begin{align}
X^{\rm H}_{\lmo} &= \left\{
\begin{array}{lr}
A^{\rm out}_{\lmo} e^{i\omega r_*} + A^{\rm in}_{\lmo}  e^{-i\omega r_*} \,, & r\rightarrow \infty\,, \\
\\
A^{\rm hole}_{\lmo} e^{-ik r_*} \,, & r\rightarrow r_H \,; 
\end{array} \right.\\ 
X^{\rm \infty}_{\lmo} &= \left\{
\begin{array}{lr}
C^{\infty}_{\lmo}  e^{i\omega r_*} \,, & r\rightarrow \infty\,, \\
\\
C^{\rm out}_{\lmo} e^{ik r_*} + C^{\rm in}_{\lmo}  e^{ -ik r_*} \,, & r\rightarrow r_H \,.
\end{array} \right.
\end{align}
The amplitudes $A$ and $C$ can be related to the amplitudes $B$ and $D$ by matching the asymptotic solutions to the SN and the Teukolsky equation on the horizon and at infinity.  The $B$-coefficients and $A$-coefficients are related by
\begin{align}
\label{eq:B-to-A-1}
B^{\rm in}_{\lmo}  &= -\frac{1}{4\omega^2} A^{\rm in}_{\lmo} \,, \\ 
\label{eq:B-to-A-2}
B^{\rm out}_{\lmo} &= -\frac{4\omega^2}{c_0} A^{\rm out}_{\lmo} \,, \\
\label{eq:B-to-A-3}
B^{\rm hole}_{\lmo} &= \frac{1}{d_{\lmo}} A^{\rm hole}_{\lmo}\,,
\end{align}
and the $D$-coefficients and $C$-coefficients are related by
\begin{align}
\label{eq:D-to-C-1}
D^{\rm in}_{\lmo}  &= \frac{1}{d_{\lmo}} C^{\rm in}_{\lmo} \,, \\ 
\label{eq:D-to-C-2}
D^{\rm out}_{\lmo} &= f_{\lmo} C^{\rm out}_{\lmo} \,, \\
\label{eq:D-to-C-3}
D^{\infty}_{\lmo} &= -\frac{4\omega^2}{c_0} C^{\infty}_{\lmo}\,,
\end{align}
where 
\begin{align}
d_{\lmo} = \sqrt{2 r_H} [ &(8-24i\omega -16\omega^2) r^2_H \\ \nn
& +  (12iam-16+16am\omega+24i\omega)r_H \\ \nn
& - 4a^2m^2-12iam+8]\,, \\
\end{align}
and 
\begin{align} 
f_{\lmo} = - \frac{4k\sqrt{2r_H} \lrsbrk{2kr_H + i(r_H-1)}}{\eta(r_H)} \,.
\end{align}
Here $\eta(r)$ is defined by
\begin{align}
\eta(r) = c_0 + \frac{c_1}{r} + \frac{c_2}{r^2} + \frac{c_3}{r^3} + \frac{c_4}{r^4}\,, \\ \nn
\end{align}
with
\begin{align}
c_0 &= -12i\omega + \lambda(\lambda+2) -12 a\omega(a\omega-m) \,, \\ \nn
c_1 &= 8ia[3a\omega - \lambda(a\omega-m)]\,, \\  
c_2 &= -24ia(a\omega-m)+ 12a^2[1-2(a\omega-m)^2]\,, \\  \nn
c_3 &= 24ia^3(a\omega-m) -24a^2\,, \\\nn
c_4 &= 12 a^4\,, \nn
\end{align}
where $\lambda \equiv {}_{-2}\lambda_{\lmo}$ is the eigenvalue of the $s=-2$ spin-weighted spheroidal harmonic.

\section{Conservation of energy for gravitational perturbations}
\label{app:conservation-erg}

In this section, we derive a new conservation relation among four energies, which corresponds to waves that are outgoing at infinity, ingoing at infinity, coming down to the ``horizon'', and being reflected from the ``horizon'', respectively.  A derivation has been done by Teukolsky and Press in~\cite{Teukolsky:1974yv} for the relation among the first three energies.  Here we extend their results to include the reflected one.

From the Newman-Penrose equations, one can derive the Teukolsky-Starobinsky identities for $s=\pm2$, which can be written as
\begin{align}
\label{eq:TS-1-1}
\mathscr{L}_{-1}\mathscr{L}_{0}\mathscr{L}_{1}\mathscr{L}_{2}\, {}_{2}S+12 i\omega {}_{-2}S & = C {}_{-2}S  \,, \\
\label{eq:TS-1-2}
\cD \cD \cD \cD {}_{-2}R            & = \frac{1}{4} \, {}_{2}R\,,
\end{align}
where we have omitted $(\lmo)$-indices in $R$ and $S$ for the sake of brevity.  We will adopt these abbreviated notations throughout this section.  The operators $\mathscr{L}$ and $\cD$ are defined by
\begin{align}
\mathscr{L}_n &= \partial_\theta + m \csc\theta -a\omega\sin\theta+ n\cot\theta \,, \\
\cD &= \partial_r - iK/\Delta \,,
\end{align}
and $C$ is given by
\begin{align}
 \vert C&\vert ^2
 = \lrbrk{(\lambda +2)^2 + 4a\omega m - 4a^2 \omega^2} \\ \nn
                & \times \lrsbrk{\lambda^2+36a\omega m -36 a^2 \omega^2} \\ \nn
                & + (2\lambda +3)(96a^2\omega^2 - 48a\omega m) + 144\omega^2(1-a^2)\,,
 \end{align} 
 with 
 \begin{align}
 \text{Im} \, C &= 12\omega\,, \\ 
 \text{Re} \, C &= +\sqrt{\vert C\vert^2-(\text{Im} \, C)^2} \,. 
 \end{align}
Similarly we define
\begin{align}
\mathscr{L}^\dagger_n & = \mathscr{L}_n (-\omega, -m) \,,\\ 
\cDD &= \cD(-\omega,-m) =\partial_r + iK/\Delta \,.
\end{align}
A complementary set of equations to Eqs.~\eqref{eq:TS-1-1} and ~\eqref{eq:TS-1-2} then gives
\begin{align}
\label{eq:TS-2-1}
\mathscr{L}^\dagger_{-1}\mathscr{L}^\dagger_{0}\mathscr{L}^\dagger_{1}\mathscr{L}^\dagger_{2} C{}_{-2}S & +12 i\omega C^* {}_{2}S  = C^2 {}_{2}S  \,, \\
\label{eq:TS-2-2}
\cDD \cDD \cDD \cDD \Delta^2 \, {}_{2}R & \!= \!4\vert C \vert^2 \Delta^{-2} {}_{-2}R \,.
\end{align}

Now let us derive the relation between $\psi_0$ and $\psi_4$ by using the Teukolsky-Starobinsky identities. Note that at large $r$, the radial function ${}_{s}R$ has the following asymptotic behavior,
\begin{align}
\label{eq:R2-inf}
{}_{2}R  &= Y_{\rm in} \frac{e^{-i\omega r_*}}{r} + Y_{\rm out} \frac{e^{i\omega r_*}}{r^5} \,, \\ 
\label{eq:Rm2-inf}
{}_{-2}R &=Z_{\rm in}\frac{e^{-i\omega r_*}}{r} + Z_{\rm out} r^3 e^{i\omega r_* }\,.
\end{align}
Plugging these asymptotic expressions into Eq.~\eqref{eq:TS-1-2}, and keep the terms leading in $(1/r)$-expansions, we have
\begin{equation}
\label{eq:Yin-Zin-relation}
CY_{\rm in} = 64\omega^4 Z_{\rm in}\,.
\end{equation}
A set of useful identities can be used during the derivations are 
\begin{align}
\Delta \cD \cD = & \, 2(-i K+r-1) \cD + 6 i\omega r+\lambda \,, \\ 
\Delta^2 \cD \cD \cD = 
 & \lrsbrk{4iK(i K-r+1) + (\lambda + 2 + 2i \omega r)\Delta} \cD  \\ \nn
 &-2iK(\lambda+6i \omega r) + 6 i \omega \Delta \,, \\
\Delta^3 \cD \cD \cD \cD = 
 & \, \left[\Delta  \lrbrk{-4 i K(\lambda +2) -8 i\omega r (r-1)  } \right. \\ \nn
 &  + \left. 8 i K \left(K^2+(r-1)^2\right) +8 i \omega  \Delta^2 \right] \cD  \\ \nn
 &  + \Delta \left[(\lambda+2 -2 i\omega r) (\lambda +6 i r \omega ) \right. \\ \nn
 & \left.-12 i \omega  (i K+r-1) \right]\\ \nn
 &  +  4 i K (i K+r-1) (\lambda +6 i r \omega ) \,.
\end{align}
Similarly, plugging the asymptotic expressions of the radial functions ${}_{\pm 2}R$ into Eq.~\eqref{eq:TS-2-2}, we obtain that
\begin{equation}
\label{eq:Yout-Zout-relation}
4\omega^4 Y_{\rm out} = C^* Z_{\rm out} \,.
\end{equation}
On the horizon, the radial function ${}_{s}R$ is given by
\begin{align}
\label{eq:R2-h}
{}_{2}R   &= Y_{\rm hole} \Delta^{-2} e^{-ikr_*} + Y_{\rm refl} e^{ikr_*}\,, \\
\label{eq:Rm2-h}
{}_{-2}R  &= Z_{\rm hole} \Delta^{+2} e^{-ikr_*} + Z_{\rm refl} e^{ikr_*}\,.
\end{align}
Plugging these expressions into Eq.~\eqref{eq:TS-1-2} and~\eqref{eq:TS-2-2}, we obtain that
\begin{align}
\label{eq:Yhole-Zhole-relation}
\!\!CY_{\rm hole} = 64 (2r_H)^4(ik)(-ik+4\epsilon)(k^2+4\epsilon^2)Z_{\rm hole} \,, \\ 
\label{eq:Yref-Zref-relation}
4(2r_H)^4 (ik)(-ik-4\epsilon)(k^2+4\epsilon^2)Y_{\rm refl} = C^* Z_{\rm refl}\,.
\end{align}

In the Schwarzschild case, the energy conservation relations can be most easily seen from the Wronskian of two linearly independent homogeneous solutions to the perturbation equations such as the Regge-Wheeler equation.   In the Teukolsky equation, due to the existence of the $dR/dr_*$-term, the Wronskian is then dependent on $r$.  To resolve this, one can rewrite the radial Teukolsky equation~\eqref{eq:Teukolsky-radial} in a form of
\begin{equation}
\label{eq:New-Teukolsky}
d^2Y/dr^2_* + V Y =0 \,,
\end{equation}
which is possible if one defines
\begin{widetext}
\begin{align}
Y &= \Delta^{s/2} (r^2+a^2)^{1/2} R \,, \\ 
V &= \frac{\lrsbrk{K^2-2isK(r-1)+\Delta(4ir\omega s-\lambda-2)-s^2(1-a^2)}}{(r^2+a^2)^2} - \frac{\Delta(2r^3+a^2r^2-4ra^2+a^4)}{(r^2+a^2)^4}\,.
\end{align}
The Wronskian of any two solutions of Eq.~\eqref{eq:New-Teukolsky} is then conserved.  By equating the Wronskian evaluated at infinity and that on the horizon, we have
\begin{equation}
\lrbrk{\frac{d {}_sY}{dr_*}{}_{-s}Y^* - {}_sY  \frac{d {}_{-s}Y^*}{dr_*}}_{r=r_H}  = \lrbrk{\frac{d{}_sY}{dr_*}{}_{-s}Y^* - {}_sY \frac{d {}_{-s}Y^*}{dr_*}}_{r=\infty}\,.
\end{equation}
For $s=2$, we substitute Eqs.~\eqref{eq:R2-inf},~\eqref{eq:Rm2-inf},~\eqref{eq:R2-h} and~\eqref{eq:Rm2-h} into the Wronskian equation, and we use Eqs.~\eqref{eq:Yin-Zin-relation},~\eqref{eq:Yout-Zout-relation},~\eqref{eq:Yhole-Zhole-relation} and~\eqref{eq:Yref-Zref-relation} to obtain that
\begin{equation}
\label{eq:erg-conserve-explicit}
\frac{-iC^* \vert Y_{\rm hole}\vert^2}{32k(2r_H)^3(k^2+4\epsilon^2)} + 
\frac{256(ik)r^5_H(k^2+4\epsilon^2)(k^2+16\epsilon^2)\vert Y_{\rm refl}\vert^2}{C} = 
\frac{-iC^* \vert Y_{\rm in}\vert^2}{32\omega^3}+
\frac{8i\omega^5\vert Y_{\rm out}\vert^2}{C}\,,
\end{equation}
where $\epsilon$ is defined in Eq.~\eqref{eq:epsilon-def}.

This is indeed the energy conservation law relating the ingoing energy at infinity $E_{\rm in}$, the outgoing energy at infinity $E_{\rm out}$, the energy absorbed by the ``horizon'' $E_{\rm hole}$, and the energy reflected from the horizon $E_{\rm refl}$.  The conservation law can be written as 
\begin{equation}
\frac{d E_{\rm in}}{d\omega}- \frac{d E_{\rm out}}{d\omega} = \frac{dE_{\rm hole}}{d\omega}-\frac{dE_{\rm refl}}{d\omega}\,,
\end{equation}
in which the explicit expressions for the four energies are
\begin{align}
\frac{d E_{\rm in}}{d\omega} &= \sum_{\lm} \frac{1}{64\pi \omega^2} \vert Y_{\rm in} \vert^2 = \sum_{\lm} \frac{64\omega^6}{\pi \vert C\vert^2}\vert Z_{\rm in}\vert^2  \,, \\
\frac{d E_{\rm out}}{d\omega} & = \sum_{\lm} \frac{1}{4\pi\omega^2}\vert Z_{\rm out}\vert^2 = \sum_{\lm} \frac{4\omega^6}{\pi \vert C \vert^2}\vert Y_{\rm in} \vert^2  \,, \\
\frac{d E_{\rm hole}}{d\omega} & = \sum_{\lm} \frac{\omega}{64\pi k(k^2+4\epsilon^2) (2r_H)^3} \vert Y_{\rm hole} \vert^2 \\
                          & = \sum_{\lm} \frac{64\omega k (k^2+4\epsilon^2)(k^2+16\epsilon^2) (2r_H)^5 }{\pi \vert C\vert^2} \vert Z_{\rm hole} \vert^2 \,, \\
\frac{d E_{\rm refl}}{d\omega} & = \sum_{\lm} \frac{\omega}{4\pi k (k^2+4\epsilon^2)(2r_H)^3} \vert Z_{\rm refl} \vert^2 \,, \\
                          & = \sum_{\lm} \frac{4\omega k  (k^2+4\epsilon^2)(k^2+16\epsilon^2)(2r_H)^5
                          }{\pi \vert C\vert^2} \vert Y_{\rm refl} \vert^2 \,.
\end{align}

\end{widetext}


\bibliographystyle{apsrev4-1}
\bibliography{BHReflectivity}

\end{document}